\documentclass[runningheads]{llncs}
\usepackage{graphicx}
\usepackage{amsmath}
\usepackage{amsfonts}
\usepackage{amssymb}
\usepackage{graphicx}
\usepackage{caption}
\usepackage{subcaption}
\begin{document}
\title{Dense monocular Simultaneous Localization and Mapping by direct surfel optimization}
%
%
\author{Emanuel Trabes \and
Julio Daniel Dondo Gazzano \and
Carlos Federico Sosa P\'{a}ez}
\authorrunning{Trabes et al.}
%
\institute{Departamento de Electr\'{o}nica\\
Universidad Nacional de San Luis, Argentina
}
\maketitle              
\begin{abstract}
This work presents a novel approach for monocular dense Simultaneous Localization and Mapping. The surface to be estimated is represented as a piecewise planar surface, defined as a group of surfels each having as parameters its position and normal. These parameters are then directly estimated from the raw camera pixels measurements, by a Gauss-Newton iterative process. As far as the authors know, this is the first time this approach is used for monocular depth estimation. The representation of the surface as a group of surfels has several advantages. It allows the recovery of robust and accurate pixel depths, without the need to use a computationally demanding depth regularization schema. This has the further advantage of avoiding the use of a physically unlikely surface smoothness prior. New surfels can be correctly initialized from the information present in nearby surfels, avoiding also the need to use an expensive initialization routine commonly needed in Gauss-Newton methods. The method was written in the GLSL shading language, allowing the usage of GPGPU thus achieving real-time. The method was tested against several datasets, showing both its depth and normal estimation correctness, and its scene reconstruction quality. The results presented here showcase the usefulness of the more physically grounded piecewise planar scene depth prior, instead of the more commonly pixel depth independence and smoothness prior.
\keywords{Depth Estimation \and Visual Odometry \and SLAM.}
\end{abstract}
\section{Introduction}

One of the most fundamental abilities that unmanned vehicles must posses is the capacity for the estimation of its relative position within the exploration area. Furthermore, many types of missions like resources detection and obstacle avoidance, require that the vehicle must posses a detailed map of its surrounding. One way to acquire such knowledge is by utilizing mapping methods, were a model of the nearby terrain is recovered from the measurements of the on-board sensors. It's well known in the literature that the problem of positioning and mapping are interconnected, and both must be solved simultaneously \cite{Grisetti_et_all_2010}.

Systems that are capable of position estimation and mapping are known in the literatura as Simultaneous Localization and Mapping (SLAM) \cite{Cadena_et_all_2016}. There is a wide choice in the type of sensor that can be utilized to accomplish the SLAM task. A very interesting choice is the monocular camera. This type sensor can have, with the current state of technology, very large resolutions and also very high refresh rates.

There are many SLAM systems currently present in the literature that successfully use a monocular camera as the only sensor for resolving the SLAM problem. One example is LSD-SLAM \cite{Engel_et_al_2014}, which uses an epipolar search approach together with a kalman filter for estimating the depth of some of the pixels of a frame taken from a position of reference. The pixels that, due to lack of texture in the frame, cannot be estimated accurately are detected and discarded, and its depths are not estimated. With the depth of the pixels and the position of the reference frame a local map is constructed. in case of the frames taken in nearby positions, the information obtained from their relative position within the local map is used to improve the reference frame pixels depth estimation. 

Another example is DSO \cite{Engel-et-al-pami2018}, which takes an even more restrictive approach with the selection of pixels to be used for depth estimation. Only a very small percentage of the pixels in the frame are used for depth estimation. The resulting map is a very sparse one, but the results in localizations are outstanding. Other works like ORB-SLAM \cite{Mur-Artal_et_al_2015} use a pre-processing step on the captured frames, to identify the areas where robust depth estimation is more plausible. This pre-processing utilizes a salience detector to filter the zones of the frame that are not well posed for depth and pose estimation. All of these methods do not try to recover a detailed map of the surrounding terrain, focusing its attention on correct position estimation. The usefulness of these methods in missions that require a detailed knowledge of the surrounding terrain is questionable.

Other approaches, like DTAM \cite{Newcombe_et_al_2011a} do try to estimate depth in every pixel. A brute force plane sweeping approach \cite{Hosni_et_al_2013} in conjunction with a regularizer are used to accomplish this task. The system is still based on a pixel-wise epipolar search. And so, the estimation in areas in the frame where there is no texture is heavily regularized with the depth of nearby pixels, in order to recover somewhat correct depth estimations.
REMODE \cite{Pizzoli_et_al_2014} uses an epipolar depth search together with a probabilistic measurement merging, again with a regularization at the end of the process. Zones in the frame where there is no texture must be, again, heavily regularized.
The work of \cite{Zienkiewicz_et_al_2016} tries to estimate a mesh from the frames taken from a monocular camera. The system utilizes first a plane sweeping method similar to \cite{Newcombe_et_al_2011a} to estimate the depth, and then this information is used to optimize every vertex of the mesh. The use of a pixel-wise estimation method makes again the zones without texture difficult to estimate and it requires the use of a depth regularizer.

It has been lately recognized in the literature that the depth smoothness hypotesis introduces a prior knowledge of the scene that many times is not accurate \cite{Engel-et-al-pami2018} \cite{Goodfellow-et-al-2016}. A better prior hypothesis on the depth of the scene is desirable.

A surfel representation for the depth estimation could potentially resolve these issues. This would replace the depth smoothness prior with a more plausible piecewise planar depth prior, preserving strong discontinuities commonly found in man-made structures. Also, the representation of a part of the scene with the same paremeters could avoid the need for a regularizer. Many works found in the literature have utilized surfels to represent the surface of the scene, but always using depth camera sensor like \cite{Wang_et_al_2019} \cite{Whelan_et_al_2016}\cite{Yan_et_al_2017}.

This work presents a dense monocular SLAM system that does not utilizes the depth smoothness prior. This method accomplishes the map estimation task by modeling the map, not as a group of independent points, but as a group of surfels. This takes advantage of the knowledge that the depth of most scenes can be modeled as a piecewise planar surface, instead of just a group of independent pixels. Every piecewise portion of the scene can be estimated with a surfel. This approach allows to directly recover robust and accurate pixel depths, even in zones where there is no good texture observation. Furthermore, this approach avoids the need for a regularization step. This way we can save on computational resources and, more important, not add a physically unlikely smoothness prior. New surfels can be initialized with the information of neighbor surfels. The entire system was written in the GLSL shading language, thus using the onboard GPU to achieve real-time. Results showing the performance of the methods are provided, demonstrating its scene reconstruction quality.


\section{Proposed Approach}

\subsection{Surfel Parameterization}
\label{sec:surfel_parameterization}

Each surfel position $p_{s}$ and normal $n_{s}$ are defined relative to the pose $P_{kf} \in SE(3)$ of a reference frame $F_{kf}$. The surfels have $3$ free parameters, $2$ belonging to the degrees of freedom of its normal $n_{s}$, and the other given by the inverse depth $id_{s}$ of the ray $r_{s}$ going from the origin of the reference frame to $p_{s}$, whose coordinates are given by

\begin{align}
	p_{s} = r_{s}/id_{s} 
	\label{point_equation}
\end{align}
 
The radius of the surfel $r$ is selected so that all surfels have the same frame-space area in the reference frame $F_{kf}$, regardless of its normal or depth. This is done so to assure the same quantity of pixel observation for every surfel, this way allowing to gather enough information for the Gauss-Newton estimation process.

Along with the aforementioned parameters, the last residual of each surfel and the time that the surfel was last seen and updated are stored for later use.

\subsection{Inverse Depth Calculation}
\label{sec:inverse_depth_calculation}

With the information provided by the estimated surfel parameters, the inverse depth $id_{u}$ of every pixel $u$ belonging to the reference frame $F_{kf}$ can be obtained. Every $u$ has a corresponding ray $r_{u}$ with coordinates

\begin{align}
	r_{u} = K^{-1}\pi^{-1}(u)
	\label{ray_equation}
\end{align}

where $K$ is the camera matrix, $\pi()$ is the homogenizing function $\pi([x, y, z]) = [x/z, y/z, 1]$ and $u \in R^{3}$ is a homogeneous pixel coordinate relative to $F_{kf}$.

To be in the same plane, the point $p_{s}$ and $p_{u} = r_{u}/id_{u}$ must satisfy $n_{s}(p_{s} - p_{u}) = 0$, so $id_{u}$ is computed as 

\begin{align}
	id_{u} = \frac{ (K^{-1}\pi^{-1}(u)) \cdot n_{s}}{(r_{s}/id_{s}) \cdot n_{s}} 
	\label{equ:depth_calculation}
\end{align}

The inverse depth $id_{u}$ will be calculated relative to a surfel $s$ when the distance between $u$ and the coordinates of the surfel $p_{s}$ projected into the frame $F_{kf}$, $u_{s} = \pi(K p_{s})$, is less than the radius $r$ of the surfel $s$ in screen space, that is

\begin{align}
	\left \| \pi(K p_{s}) - u \right \| < r
	\label{equ:surfel_radius}
\end{align}

The GLSL shading language is utilized for the implementation of the system, so it allows to seemingly introduce a depth buffer for oclution removal. This means that, if several surfel project the same pixel screen location, only the pixels with the closest depth to the camera are selected.

During inverse depth computation, along with the depth information the index of the generating surfel $s_{ind}$ is saved, for subsequent utilization as described below.

\subsection{Surfel Initialization}
\label{sec:surfel_initialization}

To initialize the surfels, the reference frame $F_{kf}$'s inverse depth is computed. Clearly, in areas in the image where no surfel satisfy the equation \ref{equ:surfel_radius}, there will be no inverse depth measurement. These areas are identified as potential location to initialize a new surfel. 

First, a search is carried out for an empty pixel $u_{ns}$ where a new surfel can be placed. This area must satisfy that there is no other inverse depth measurement nearby satisfying $\forall u_{nn}, \left \| u_{ns} - u_{nn} \right \| > \alpha r$, where $u_{nn}$ is a non-empty pixel.

Then, all neighboring surfels are searched for. A neighbor surfel must satisfy that a corresponding inverse depth measurement pixel $u_{s}$ is nearby, with a distance of less than
$\left \| u_{ns} - u_{s} \right \| < \beta r$

The inverse depth for the new surfel is calculated as the mean of the estimated inverse depth, following equation \ref{equ:depth_calculation} with respect of all nearby surfels $s$. The normal of the new surfel is initialized as the mean of all neighboring surfel's normal.

In so far as the real surface is globally plane, this type of initialization will give good results.
This allows to initialize the surfel with the approximate correct surfel normal and depth, which allows to initialize the Gauss-Newton optimization, without the need to search for surfel initializing parameters.

\subsection{Surfel Optimization}
\label{sec:surfel_optimization}

The parameters of the surfels are estimated so that they minimize the cost function:
\begin{equation}
	C(n_{s},id_{s}) = \sum_{n \in \Omega} \sum_{u \in s} \left \| I_{f_{n}}(u_{p}(n_{s}, id_{s})) - I_{kf}(u) \right \|_{h}
	\label{equ:cost_function}
\end{equation}

which is the commonly used sum of the huber normed photometric error \cite{Newcombe_et_al_2011a} \cite{Pizzoli_et_al_2014} \cite{Engel-et-al-pami2018} between the reference frame intensity $I_{kf}$ and every frame $I_{f_{n}}$ in a group $\Omega$ of frames taken before $I_{f_{n}}$.

The coordinates $u_{p}$ of the pixel $u$ as seen from the frame $I_{f_n}$ can be calculated with

\begin{equation}
	u_{p}(n_{s}, id_{s}) = \pi^{-1}(KP_{kf}^{f_{n}} K^{-1}\pi(u)/id_{u}(n_{s}, id_{s}))
	\label{equ:reprojection}
\end{equation}

where $P_{kf}^{f_{n}}\in SE(3)$ changes a point $p$ in the keyframe reference frame to the frame $f_{n}$ reference frame.

To recover the parameters of each surfel $[n_{s},id_{s}]$ that minimizes $C([n_{s},id_{s}])$ , a Levenber-Marquat \cite{Engel-et-al-pami2018} Gauss-Newton optimization approach is implemented. The Jacobian $\frac{\partial C_{[n_{s}, d_{s}]}}{\partial [n_{s}, d_{s}]}$ is obtained by using the chain rule for derivatives:

\begin{equation}
	J =  \sum_{n \in \Omega} \frac{\partial C([n_{s}, d_{s}])}{\partial id_{u}} \frac{\partial id_{u}}{\partial [n_{s}, d_{s}]}
	\label{equ:jacobian}
\end{equation}

$\frac{C([n_{s}, d_{s}])}{\partial id_{u}}$ can be further expressed as

\begin{equation}
	\frac{C([n_{s}, d_{s}])}{\partial id_{u}} =  \sum_{n \in \Omega} \frac{\partial C([n_{s}, d_{s}])}{\partial I_{f}} \nabla I_{f_{n}}(up) \frac{\partial \pi(u)}{\partial u} \frac{\partial u_{p}}{\partial id_{u}}
	\label{equ:cost_derivative}
\end{equation}

where $\frac{\partial C([n_{s}, d_{s}])}{\partial I_{f}} = w (I_{f_{n}}(u_{p}) -I_{kf}(u))$, w being the Huber norm coefficient correction to the squared norm,  $\nabla I_{f_{n}}$ is the gradient of the frame $f_{n}$, $\frac{\partial \pi(u)}{\partial u} = \begin{bmatrix}
1 & 0 & -u_{y} \\
0 & 1 & -u_{y}\\ 
0 & 0 & 0
\end{bmatrix}$ 

and $\frac{\partial u_{p}}{\partial id_{u}}= KR_{kf}^{f_{n}}K^{-1}\pi^{-1}(u)$, where $R_{kf}^{f_{n}}$ is the relative rotation between the coordinate frames $P_{kf}$ and $P_{f_{n}}$.

Furthermore, $\frac{\partial id_{u}}{\partial [n_{s}, d_{s}]}$ can be expressed as

\begin{equation}
\begin{gathered}
	\frac{\partial id_{u}}{\partial n_{s}} = \frac{r_{s}/id_{s}}{K^{-1}\pi^{-1}(u)\cdot n_{s}} - \frac{(r_{s}/id_{s})K^{-1}\pi^{-1}(u)\cdot n_{s}}{(K^{-1}\pi^{-1}(u)\cdot n_{s})^{2}} \\
	\frac{\partial id_{u}}{\partial id_{s}} = \frac{K^{-1}\pi^{-1}(u) \cdot n_{s}}{r_{s} \cdot n_{s}} 
\end{gathered}
	\label{equ:depth_derivative}
\end{equation}

As $\frac{\partial id_{u}}{\partial [n_{s}, d_{s}]}$ does not depend upon $I_{f_{n}}$ or its pose $P_{kf}^{f_{n}}$, it can be taken out of the sum and computed only once for all $I_{f_{n}} \in \Omega$, which allows to reduce the computational burden of the GPU.

Notice that even $n_{s}$ has two degrees of freedom, the optimization as calculated with the equation \ref{equ:depth_derivative} utilizes tree degrees of freedom for the normal. This is done to avoid such normal expression as in cylindrical parameterizations, which adds singularities to the optimization procedure that can cause problems. It was chosen to implement a 3 degree of freedom parameterizations, and then normalize the resulting updated normal.

\subsection{Surfel Reference Frame Change}
\label{sec:surfel_reference_frame_change}

A new reference frame $F_{kf_{1}}$ is selected as in the approach described in \cite{Engel_et_al_2014}. First, the new reference frame pose $P_{kf_{0}}^{kf_{1}}$ is estimated, relative to the previous reference frame $P_{kf_{0}}$. If there are surfels previously estimated in the last reference frame $F_{kf_{0}}$, its surfels are passed to the new reference frame. The position and normal of each surfel are changed as

\begin{equation}
\begin{gathered}
    p_{s_{kf_1}} = P_{kf_{0}}^{kf_{1}} p_{s_{kf_0}}\\
	n_{kf_{1}} = R_{kf_{0}}^{kf_{1}} n_{kf_{0}}\\
	\label{equ:surfel_reprojection}
\end{gathered}
\end{equation}

where $R_{kf_{0}}^{kf_{1}} \in SO(3)$ is the rotation of $P_{kf_{0}}^{kf_{1}}$. The ray $r_{s_{kf_{1}}}$ can be easily calculated as $\pi(p_{kf_{1}})$. 

\section{Implementation}

The proposed system is implemented using the parallel computing capabilities of common GPU, by utilizing OpenGL core 3.3. This way good performance is achieved without the need to have a high performance system capable of CUDA or OpenCL.

This depth estimation approch is integrated with the LSD-SLAM system \cite{Engel_et_al_2014}, thus having position estimation and map management. It was found that this system is a good choice, because of its state of the art performance while being open-source and easy to expand upon.


\section{System evaluation}

The system was evaluated in 3 datasets available in \cite{tum_lsd_slam_datasets}. These datasets depict 3 different scenes. The first dataset was taken on food court, with tables and benches in view. The street road present in the scene has a very smooth texture, so estimating its depth can be difficult. The second dataset was taken from inside a launch buffet, so the scene has many non-diffuse surfaces, as the floor and doors. The last scene was taken walking around a machine building, so there are many open spaces and intricate structures.

In the following sections, results are shown for these three datasets. The color codes present in the figures are as follow. For the normal estimations, blue represents vectors whose directions are facing to the right edge of the frame, and red represents vectors wich are pointing to the right of the frame. Similarly, green represents vectors pointing up, and yellow are vectors pointing down. For the depth estimation figures, black represents a low inverse depth, and therefore high depth. Conversely, white represents high inverse depths, and so depths close to the camera focal point.

For the following evaluations, the camera image has a $640x480$ resolution, and the surfel size was set to a radius of $10$ pixels.

\subsection{Foodcourt Dataset}

\begin{figure}[h]
     \centering
     \begin{subfigure}[b]{0.24\textwidth}
         \centering
         \includegraphics[width=0.99\textwidth, height=0.7\textwidth]{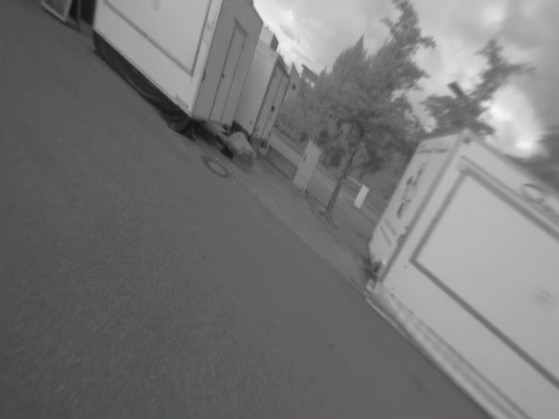}
     \end{subfigure}
     \begin{subfigure}[b]{0.24\textwidth}
         \centering
         \includegraphics[width=0.99\textwidth, height=0.7\textwidth]{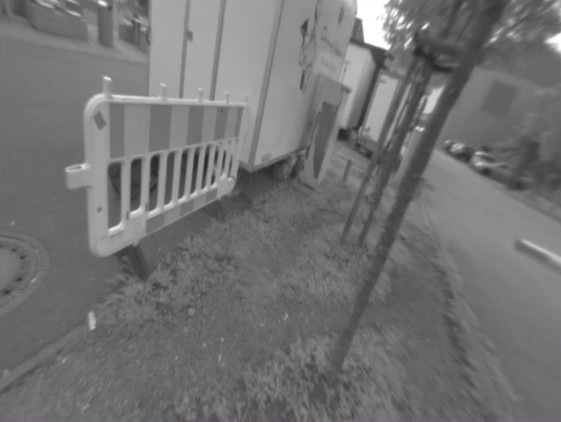}
     \end{subfigure}
     \begin{subfigure}[b]{0.24\textwidth}
         \centering
         \includegraphics[width=0.99\textwidth, height=0.7\textwidth]{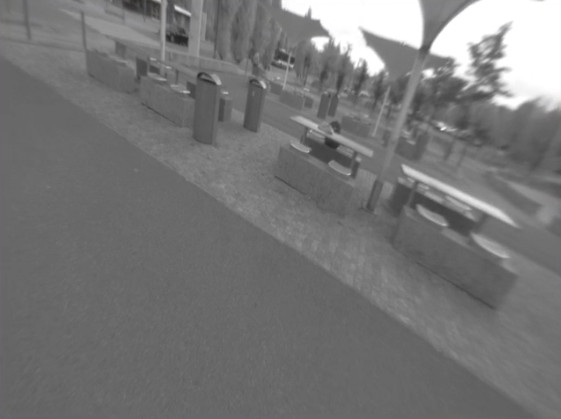}
     \end{subfigure}
     \centering
     \begin{subfigure}[b]{0.24\textwidth}
         \centering
         \includegraphics[width=0.99\textwidth, height=0.7\textwidth]{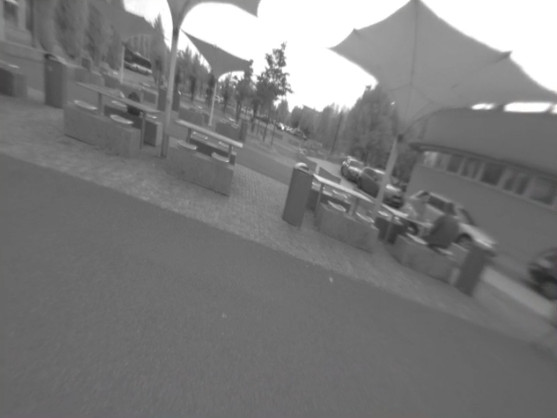}
     \end{subfigure}
     \begin{subfigure}[b]{0.24\textwidth}
         \centering
         \includegraphics[width=0.99\textwidth, height=0.7\textwidth]{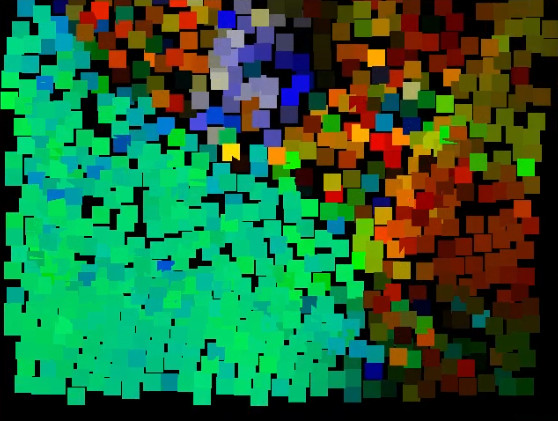}
     \end{subfigure}
     \begin{subfigure}[b]{0.24\textwidth}
         \centering
         \includegraphics[width=0.99\textwidth, height=0.7\textwidth]{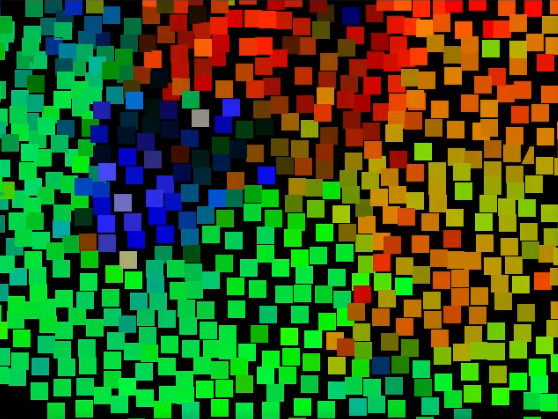}
     \end{subfigure}
     \begin{subfigure}[b]{0.24\textwidth}
         \centering
         \includegraphics[width=0.99\textwidth, height=0.7\textwidth]{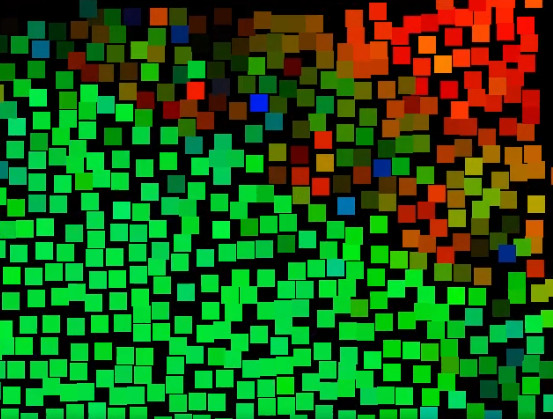}
     \end{subfigure}
     \begin{subfigure}[b]{0.24\textwidth}
         \centering
         \includegraphics[width=0.99\textwidth, height=0.7\textwidth]{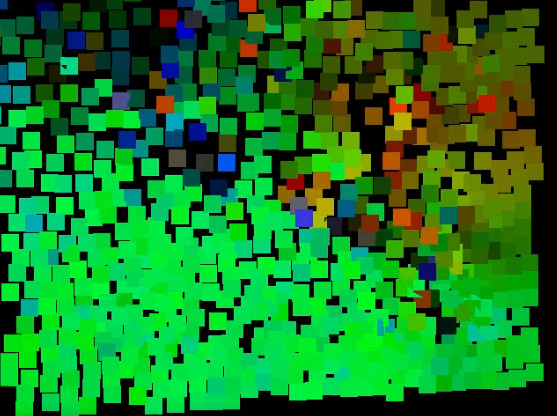}
     \end{subfigure}
     \begin{subfigure}[b]{0.24\textwidth}
         \centering
         \includegraphics[width=0.99\textwidth, height=0.7\textwidth]{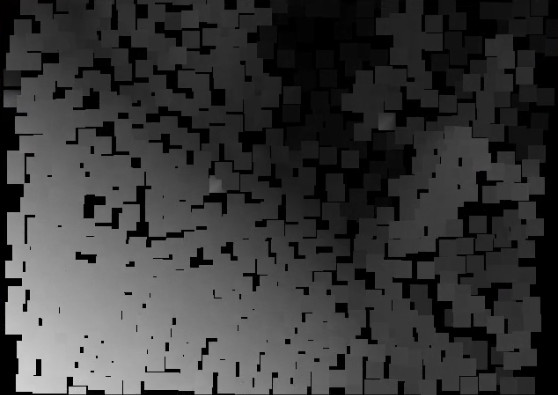}
     \end{subfigure}
     \begin{subfigure}[b]{0.24\textwidth}
         \centering
         \includegraphics[width=0.99\textwidth, height=0.7\textwidth]{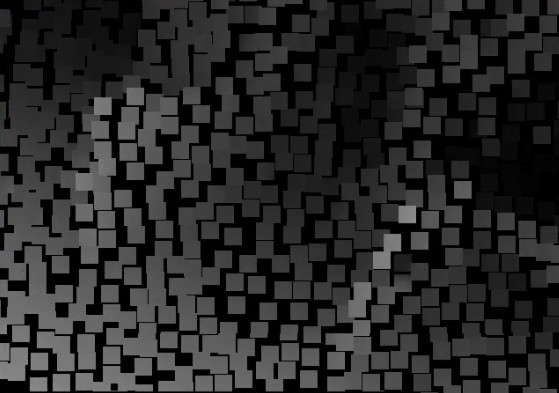}
     \end{subfigure}
     \begin{subfigure}[b]{0.24\textwidth}
         \centering
         \includegraphics[width=0.99\textwidth, height=0.7\textwidth]{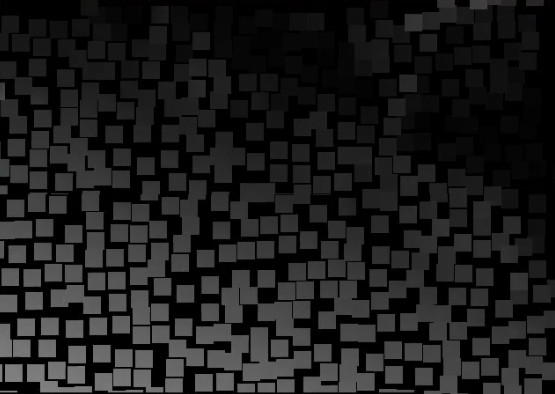}
     \end{subfigure}
     \begin{subfigure}[b]{0.24\textwidth}
         \centering
         \includegraphics[width=0.99\textwidth, height=0.7\textwidth]{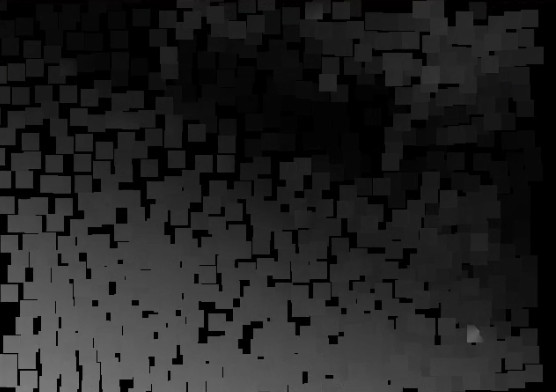}
     \end{subfigure}
        \caption{Foodcourt Dataset results. First column: raw frame. Second column: estimated normals. Third column: estimated depth.}
        \label{fig:foodcourt_dataset}
\end{figure}

Results from the Foodcourt dataset are shown in the figure \ref{fig:foodcourt_dataset}. It can be seen that the resulting depth is correctly recovered, even in the low texture sections of the street. The first column of figure \ref{fig:foodcourt_dataset} depict a backward moving movement. This means that the pixels in the lower part of the image are pixels just entering the frame, and so they have received very few observations. This highlights the usefulness of the surfel initialization scheme explained in section \ref{sec:surfel_initialization}. The third column of figure \ref{fig:foodcourt_dataset} depicts a forward moving movement, showing that both types of movements allow correct normal and depth estimations. In the second column it can be seen that the guarding fence, the food truck and the floor have correct depth and normal estimations. This part of the dataset is particularly challenging, because the guarding fence was seen a few frame ago from the opposite side, so its normals had the inverse direction. The same happens with the normals of the food truck.

\subsection{Eccv Dataset}

\begin{figure}[h]
     \centering
     \begin{subfigure}[b]{0.24\textwidth}
         \centering
         \includegraphics[width=0.99\textwidth, height=0.7\textwidth]{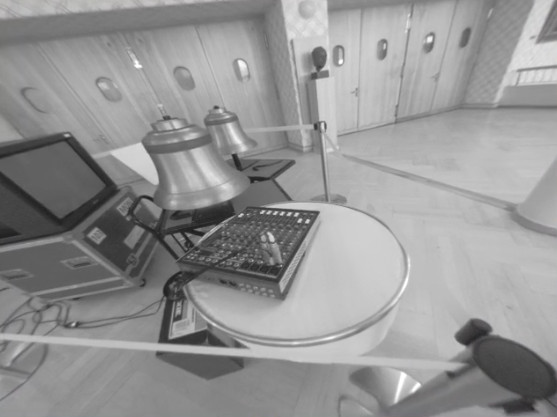}
     \end{subfigure}
     \begin{subfigure}[b]{0.24\textwidth}
         \centering
         \includegraphics[width=0.99\textwidth, height=0.7\textwidth]{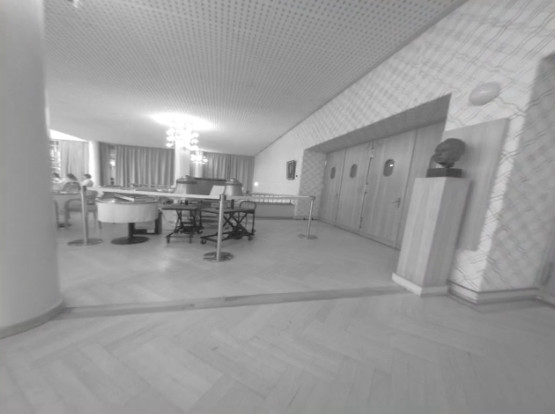}
     \end{subfigure}
     \begin{subfigure}[b]{0.24\textwidth}
         \centering
         \includegraphics[width=0.99\textwidth, height=0.7\textwidth]{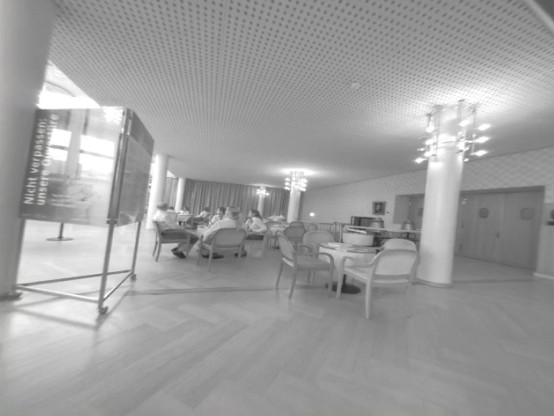}
     \end{subfigure}
     \centering
     \begin{subfigure}[b]{0.24\textwidth}
         \centering
         \includegraphics[width=0.99\textwidth, height=0.7\textwidth]{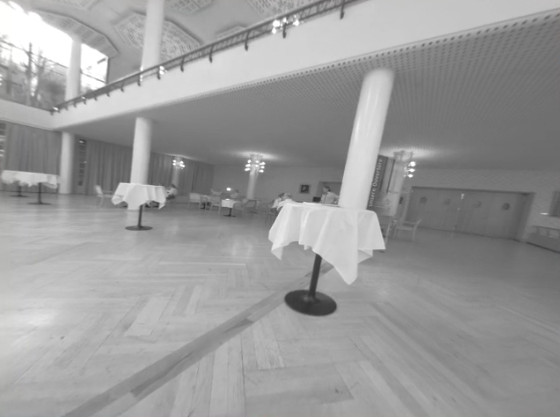}
     \end{subfigure}
     \begin{subfigure}[b]{0.24\textwidth}
         \centering
         \includegraphics[width=0.99\textwidth, height=0.7\textwidth]{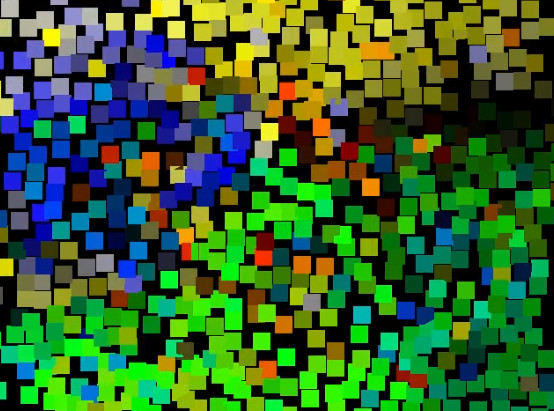}
     \end{subfigure}
     \begin{subfigure}[b]{0.24\textwidth}
         \centering
         \includegraphics[width=0.99\textwidth, height=0.7\textwidth]{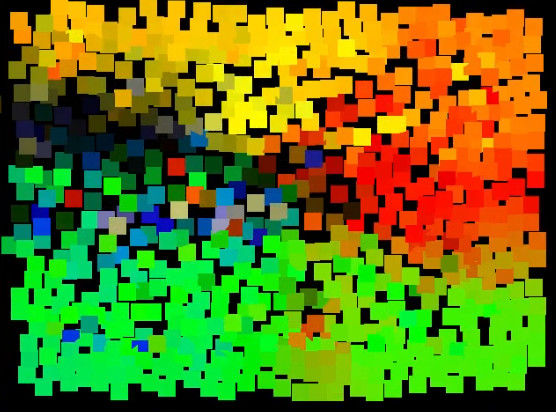}
     \end{subfigure}
     \begin{subfigure}[b]{0.24\textwidth}
         \centering
         \includegraphics[width=0.99\textwidth, height=0.7\textwidth]{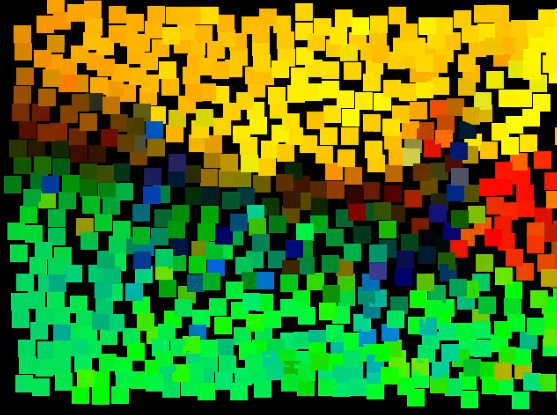}
     \end{subfigure}
     \begin{subfigure}[b]{0.24\textwidth}
         \centering
         \includegraphics[width=0.99\textwidth, height=0.7\textwidth]{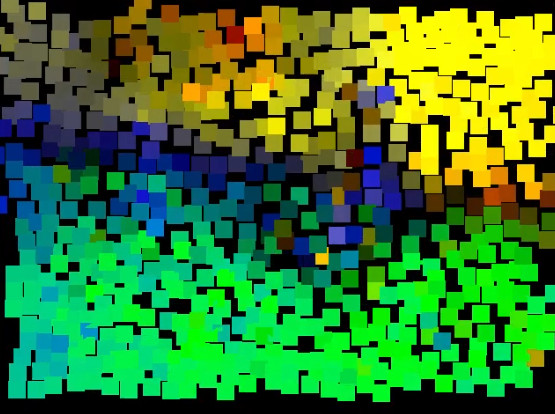}
     \end{subfigure}
     \begin{subfigure}[b]{0.24\textwidth}
         \centering
         \includegraphics[width=0.99\textwidth, height=0.7\textwidth]{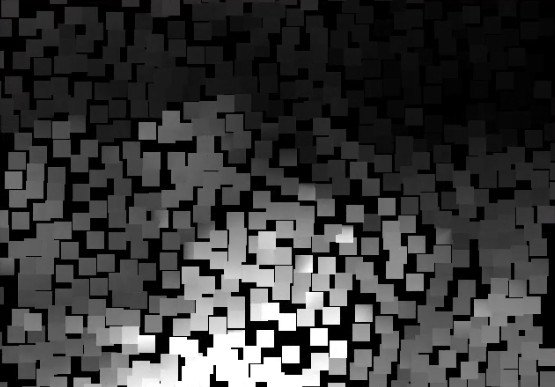}
     \end{subfigure}
     \begin{subfigure}[b]{0.24\textwidth}
         \centering
         \includegraphics[width=0.99\textwidth, height=0.7\textwidth]{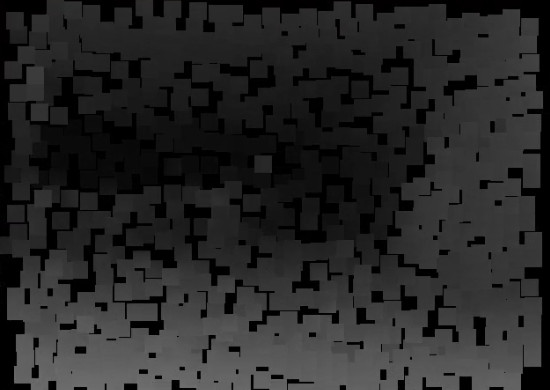}
     \end{subfigure}
     \begin{subfigure}[b]{0.24\textwidth}
         \centering
         \includegraphics[width=0.99\textwidth, height=0.7\textwidth]{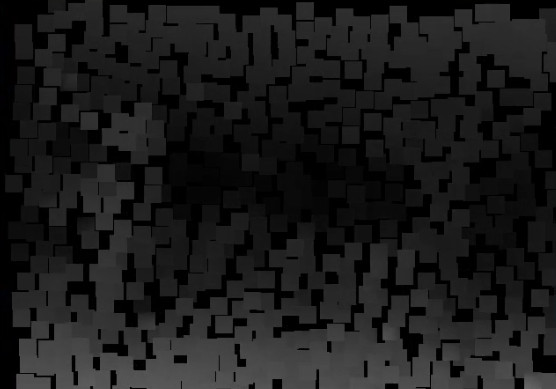}
     \end{subfigure}
     \begin{subfigure}[b]{0.24\textwidth}
         \centering
         \includegraphics[width=0.99\textwidth, height=0.7\textwidth]{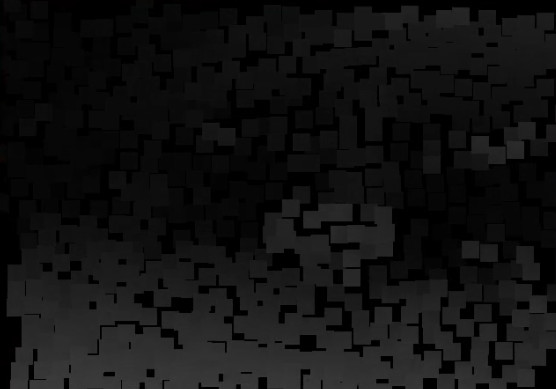}
     \end{subfigure}
        \caption{Eccv Dataset results. First column: raw frame. Second column: estimated normals. Third column: estimated depth.}
        \label{fig:eccv_dataset}
\end{figure}

This dataset is particularly challenging, because of the specularities present in almost all of the surfaces. The floor, the ceiling and even some of the furniture present in the scene have this characteristic. The first column of figure \ref{fig:eccv_dataset} shows that the floor depth and normal was correctly estimated, even when there are specularities and blurriness. In the second, third and fourth columns of figure \ref{fig:eccv_dataset} it can be seen that the depth and normal of the ceiling was correctly recovered. The movement in the second row was backward, so all of the pixels in the lower part of the frame are new and with very few observations. Again this shows the usefulness of the surfel initialization.

\subsection{Machine Dataset}

\begin{figure}[h]
     \centering
     \begin{subfigure}[b]{0.24\textwidth}
         \centering
         \includegraphics[width=0.99\textwidth, height=0.7\textwidth]{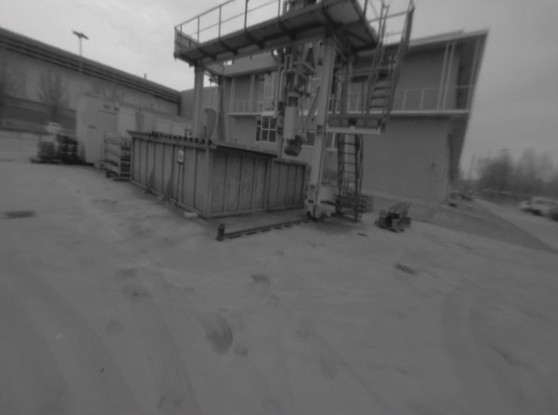}
     \end{subfigure}
     \begin{subfigure}[b]{0.24\textwidth}
         \centering
         \includegraphics[width=0.99\textwidth, height=0.7\textwidth]{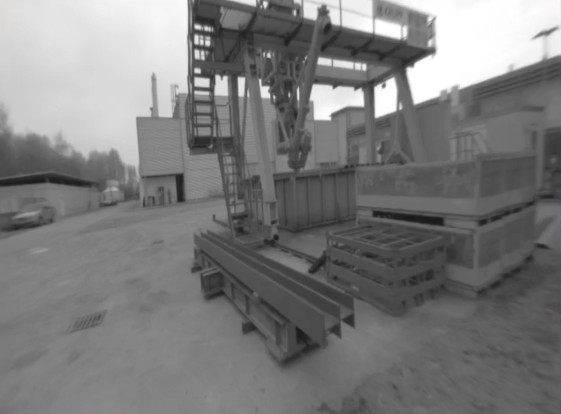}
     \end{subfigure}
     \begin{subfigure}[b]{0.24\textwidth}
         \centering
         \includegraphics[width=0.99\textwidth, height=0.7\textwidth]{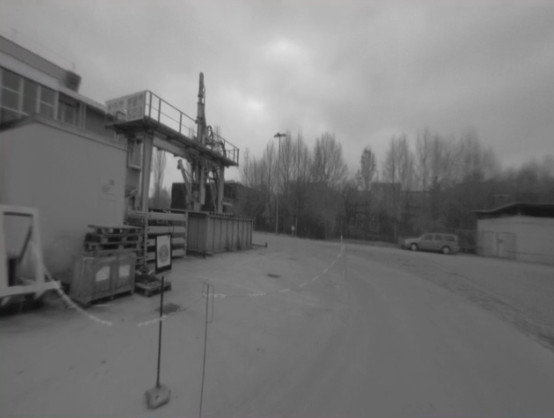}
     \end{subfigure}
     \centering
     \begin{subfigure}[b]{0.24\textwidth}
         \centering
         \includegraphics[width=0.99\textwidth, height=0.7\textwidth]{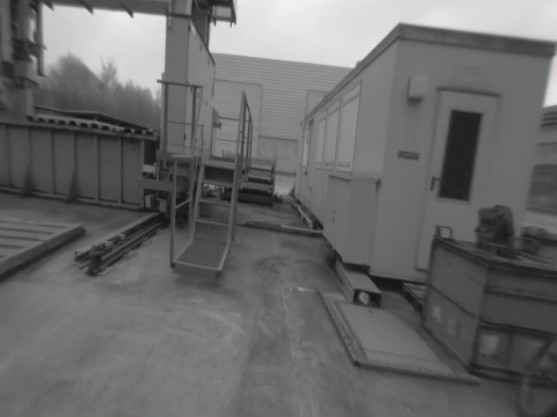}
     \end{subfigure}
     \begin{subfigure}[b]{0.24\textwidth}
         \centering
         \includegraphics[width=0.99\textwidth, height=0.7\textwidth]{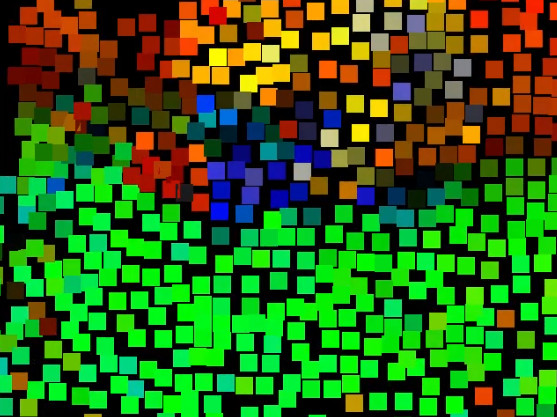}
     \end{subfigure}
     \begin{subfigure}[b]{0.24\textwidth}
         \centering
         \includegraphics[width=0.99\textwidth, height=0.7\textwidth]{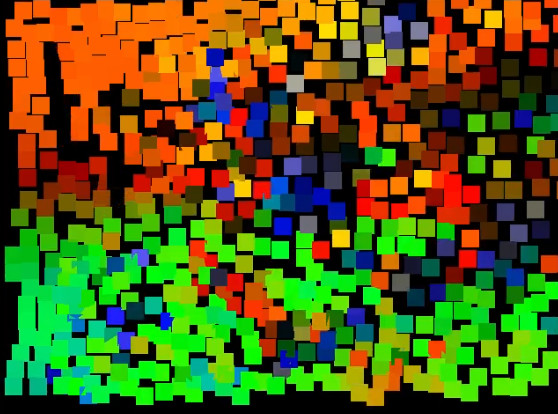}
     \end{subfigure}
     \begin{subfigure}[b]{0.24\textwidth}
         \centering
         \includegraphics[width=0.99\textwidth, height=0.7\textwidth]{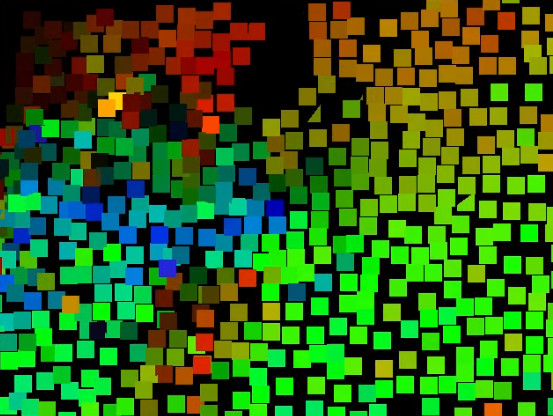}
     \end{subfigure}
     \begin{subfigure}[b]{0.24\textwidth}
         \centering
         \includegraphics[width=0.99\textwidth, height=0.7\textwidth]{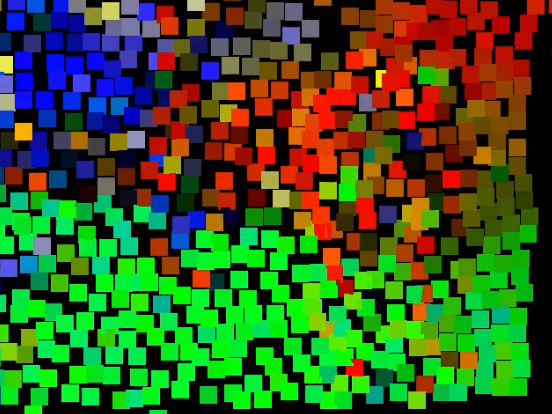}
     \end{subfigure}
     \begin{subfigure}[b]{0.24\textwidth}
         \centering
         \includegraphics[width=0.99\textwidth, height=0.7\textwidth]{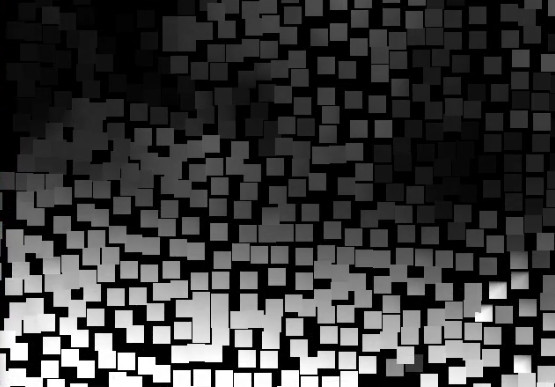}
     \end{subfigure}
     \begin{subfigure}[b]{0.24\textwidth}
         \centering
         \includegraphics[width=0.99\textwidth, height=0.7\textwidth]{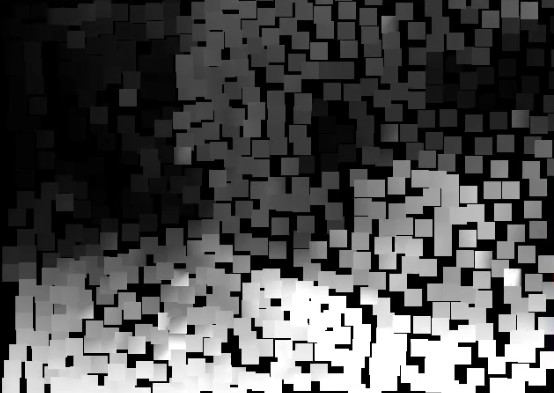}
     \end{subfigure}
     \begin{subfigure}[b]{0.24\textwidth}
         \centering
         \includegraphics[width=0.99\textwidth, height=0.7\textwidth]{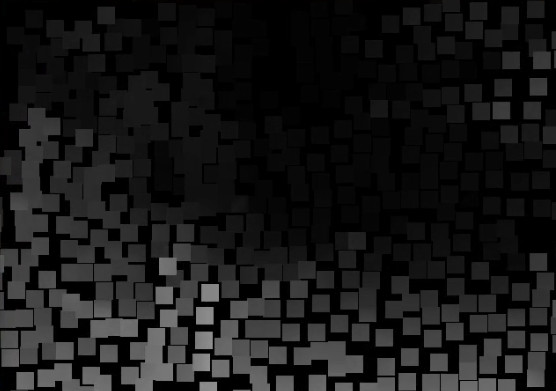}
     \end{subfigure}
     \begin{subfigure}[b]{0.24\textwidth}
         \centering
         \includegraphics[width=0.99\textwidth, height=0.7\textwidth]{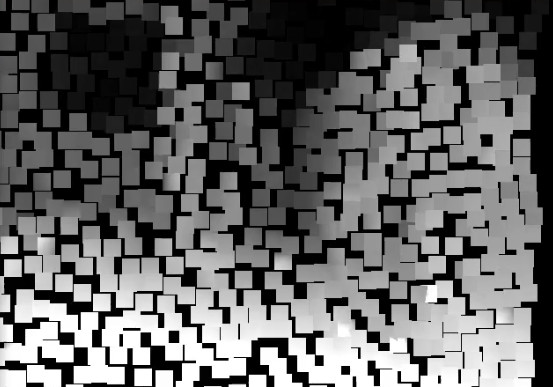}
     \end{subfigure}
     \caption{Machine Dataset results. First column: raw frame. Second column: estimated normals. Third column: estimated depth.}
             \label{fig:machine_dataset}
\end{figure}

This dataset has the particularity that it was taken outside, and the sky is a very prominent part of most of the frames. As it was a cloudy day, the sky has texture, even when it is very smooth with a very low gradient. As can be seen in figure \ref{fig:machine_dataset}, the depth of the sky was correctly recovered, even with the conditions described before.

The normals of the scene 

\subsection{Usefulness of Normal Estimation}

The usefullness of the joint depth and normal estimation was tested, by modifing the Levenberg-Marquat estimation algoritm described in \ref{sec:surfel_optimization}. In \ref{equ:depth_derivative}, $\frac{\partial id_{u}}{\partial n_{s}}$ was set to $[0.0,0.0,0.0]$, efectively just optimizing for $id_{s}$, similarly to aproches like \cite{Engel-et-al-pami2018}.

\begin{figure}
     \centering
     \begin{subfigure}[b]{0.24\textwidth}
         \centering
         \includegraphics[width=\textwidth, height=0.7\textwidth]{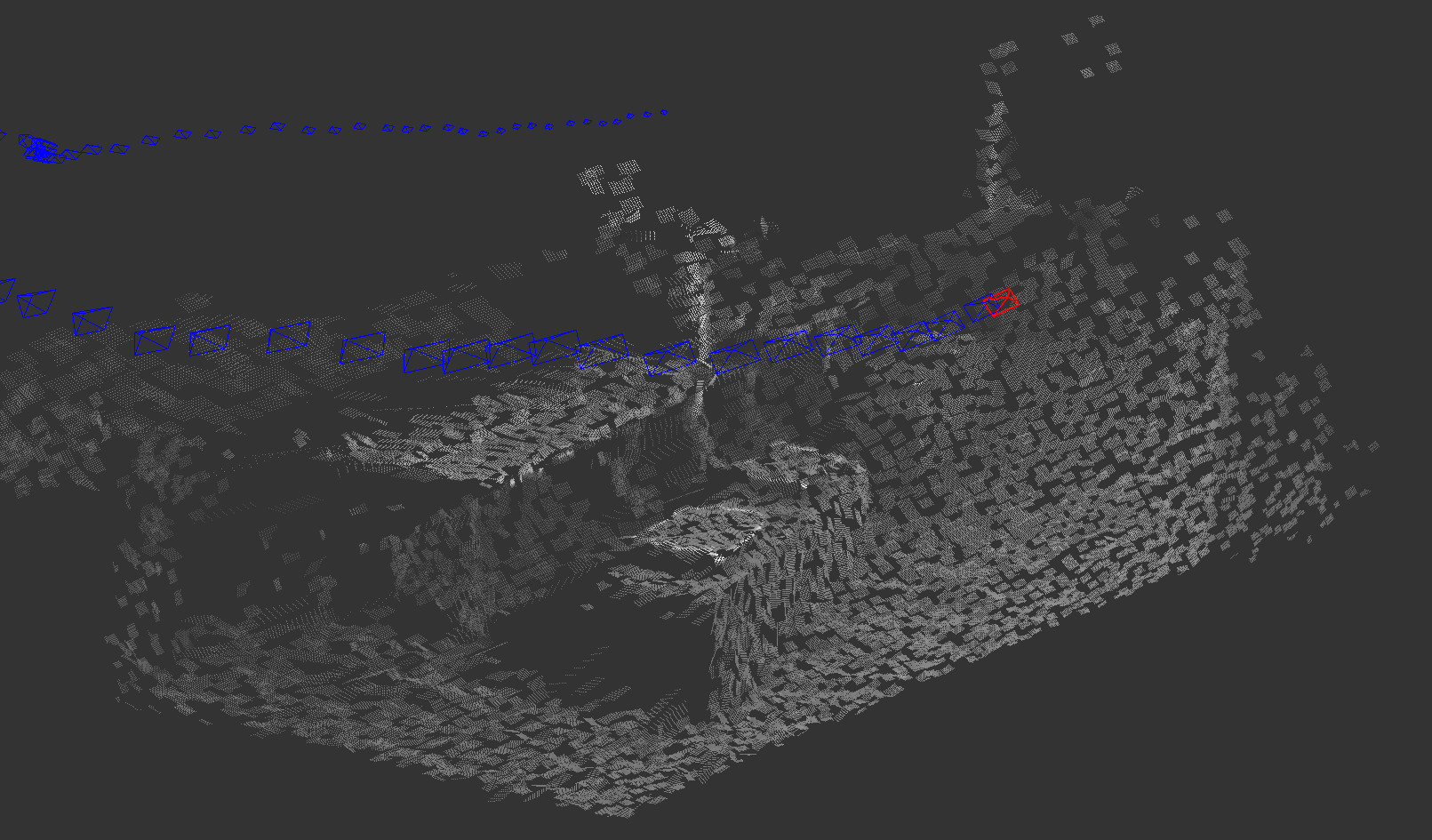}
     \end{subfigure}
     \begin{subfigure}[b]{0.24\textwidth}
         \centering
         \includegraphics[width=\textwidth, height=0.7\textwidth]{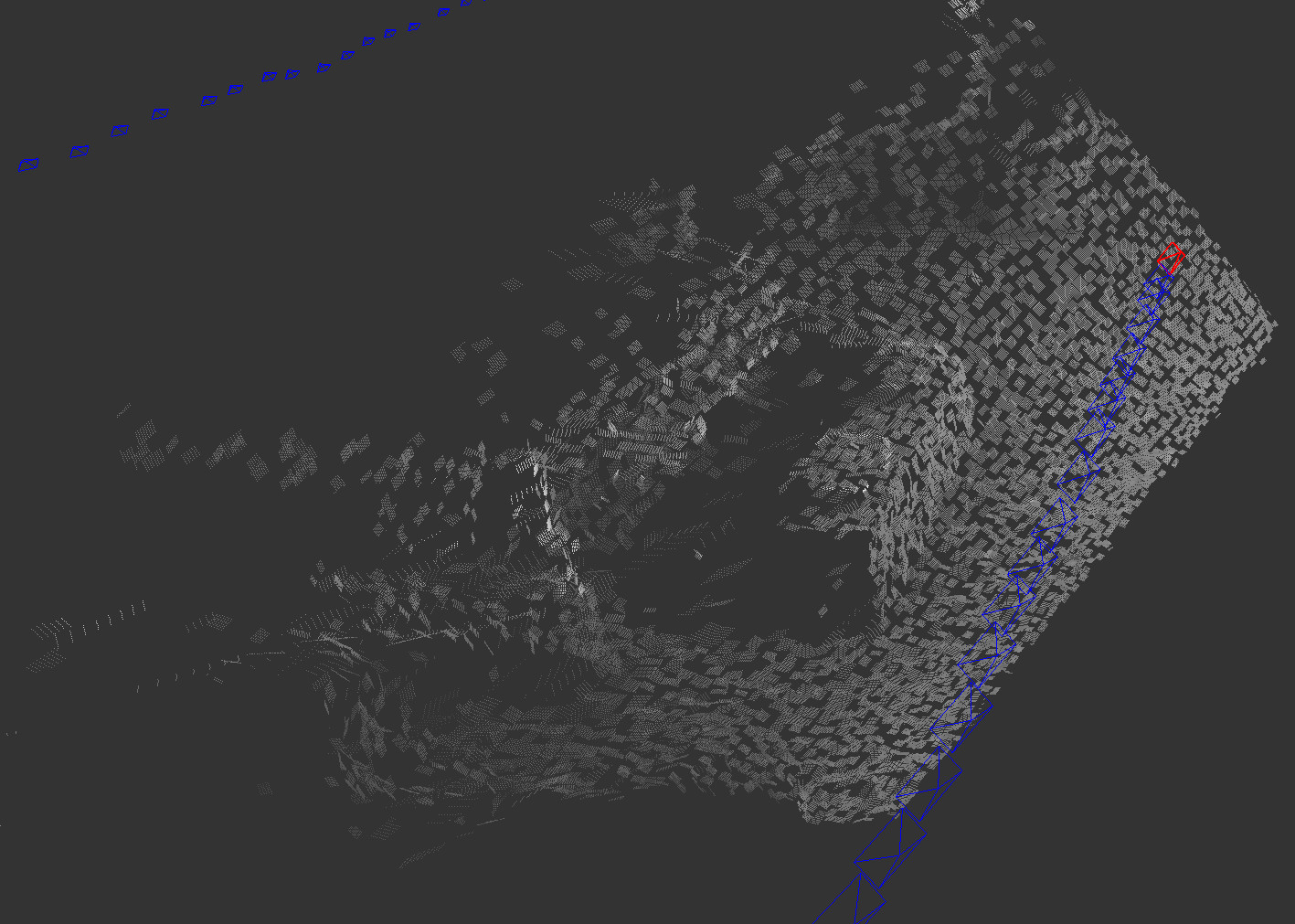}
     \end{subfigure}
     \begin{subfigure}[b]{0.24\textwidth}
         \centering
         \includegraphics[width=\textwidth, height=0.7\textwidth]{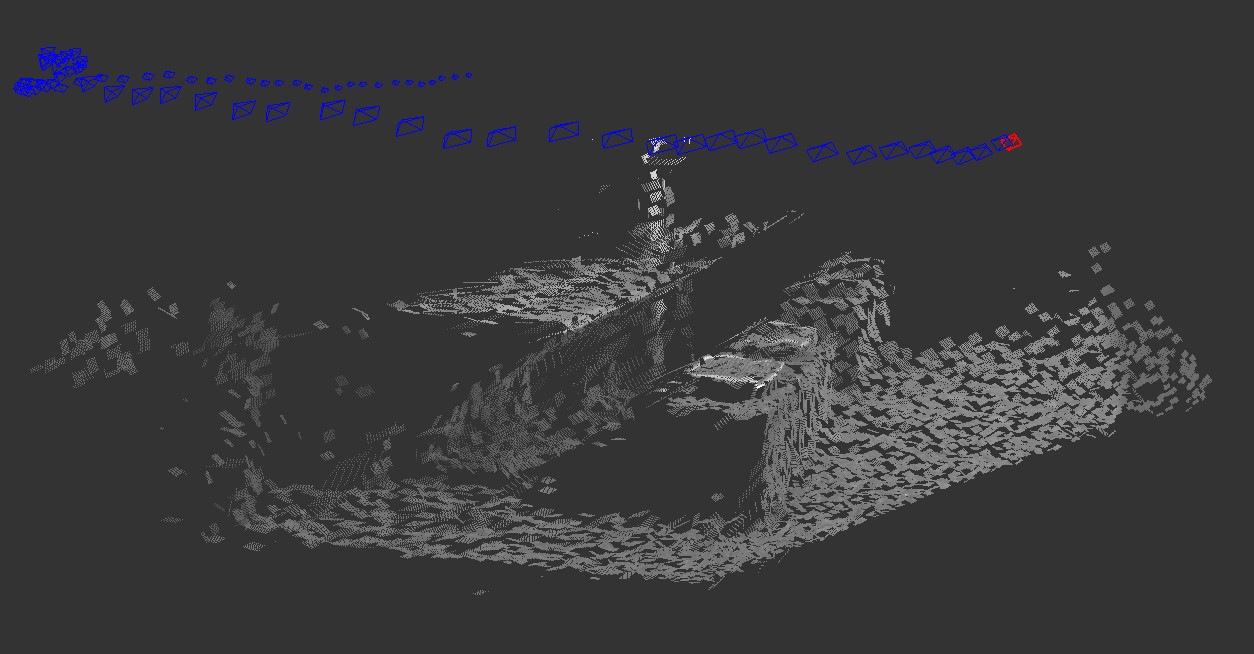}
     \end{subfigure}
     \begin{subfigure}[b]{0.24\textwidth}
         \centering
         \includegraphics[width=\textwidth, height=0.7\textwidth]{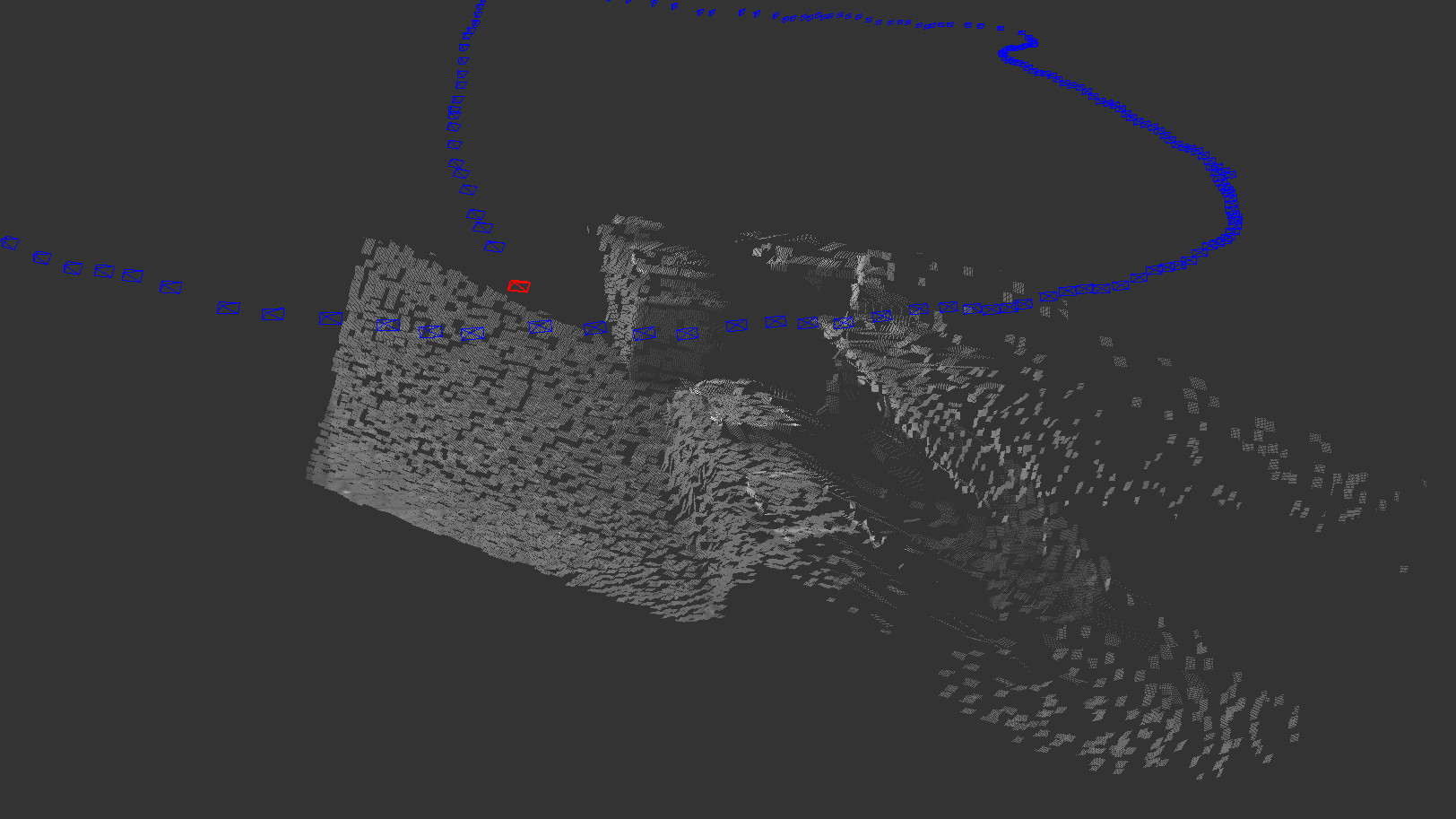}
     \end{subfigure}     
     \begin{subfigure}[b]{0.24\textwidth}
         \centering
         \includegraphics[width=\textwidth, height=0.7\textwidth]{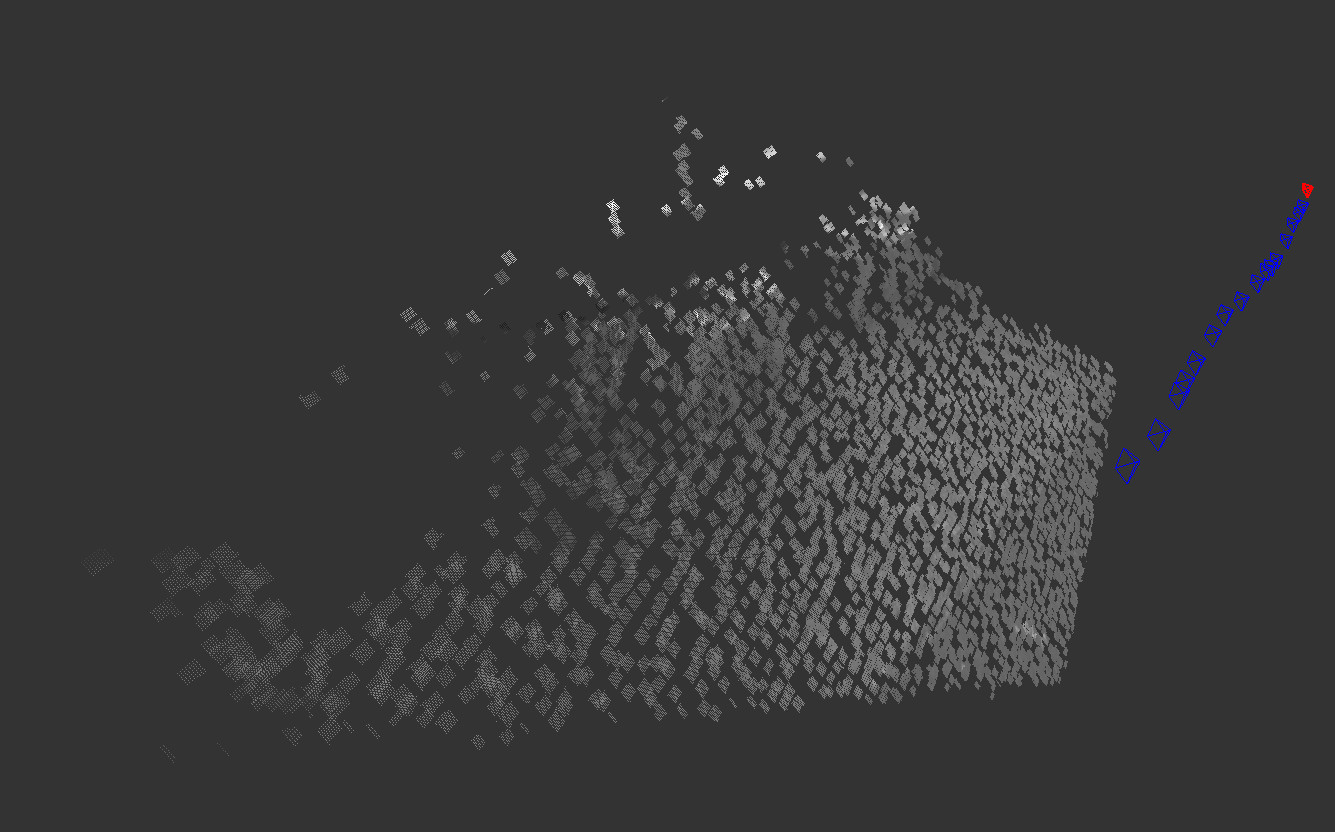}
     \end{subfigure}
     \begin{subfigure}[b]{0.24\textwidth}
         \centering
         \includegraphics[width=\textwidth, height=0.7\textwidth]{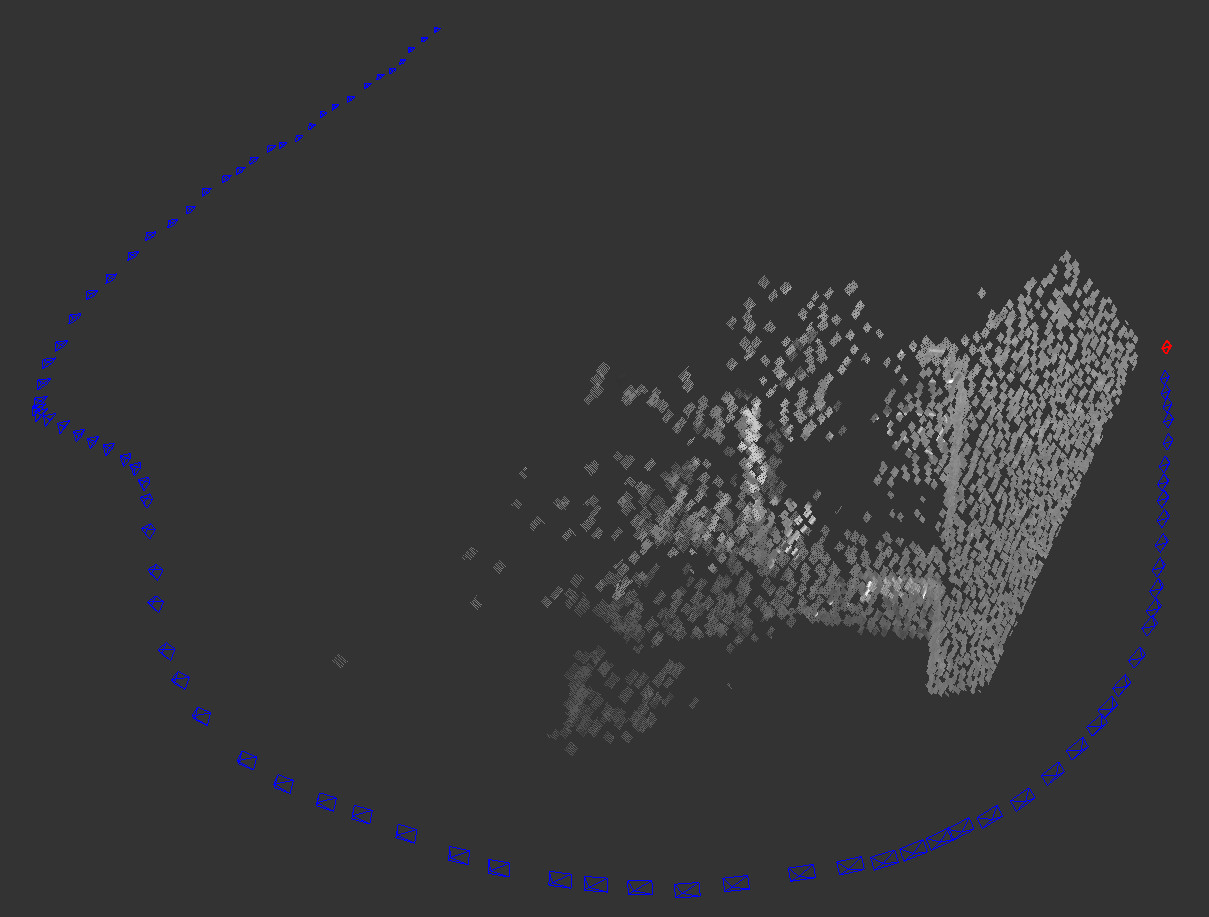}
     \end{subfigure}
     \begin{subfigure}[b]{0.24\textwidth}
         \centering
         \includegraphics[width=\textwidth, height=0.7\textwidth]{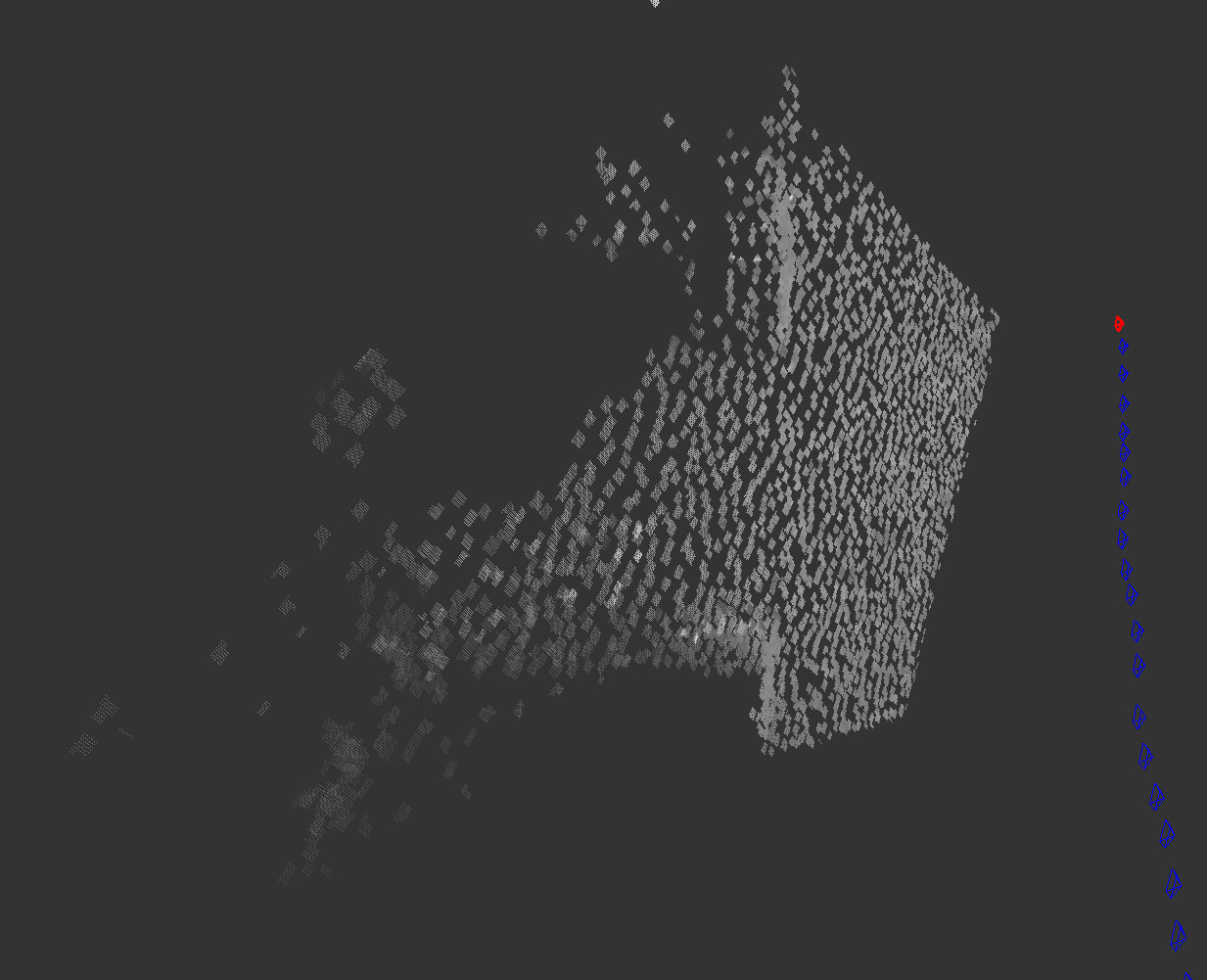}
     \end{subfigure}
     \begin{subfigure}[b]{0.24\textwidth}
         \centering
         \includegraphics[width=\textwidth, height=0.7\textwidth]{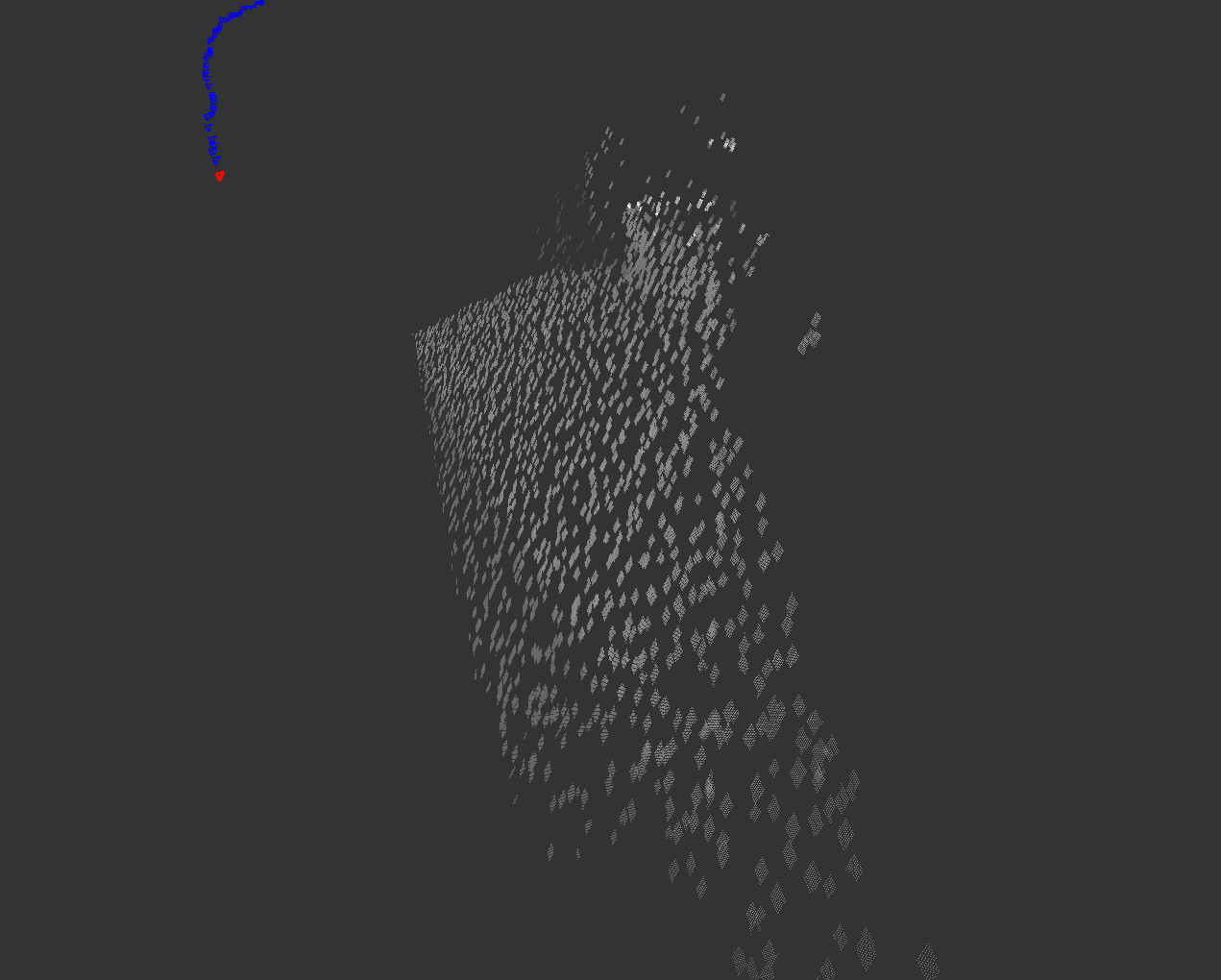}
     \end{subfigure}
     \caption{Detail of reconstruction, using a surfel radius of 5 pixels. First row: proposed method. The benches and tables present in the scene are easily recognizable, resulting from the correct surface normal estimation. Second row: without estimating surfels normals. This time the scene cannot be easily recognized, and artifacts caused by incorrect surfel normals are present}
     \label{fig:surfel_5}
\end{figure}

The resulting surface reconstruction of the foodcourt dataset with and without normal estimation can be seen in the figure \ref{fig:surfel_5}. The resulting scene reconstruction has a quality not present in the reconstruction made without normal estimation. 

\subsection{Impact of Surfel Size Selection}

\begin{figure}[h]
     \centering
     \begin{subfigure}[b]{0.24\textwidth}
         \centering
         \includegraphics[width=0.99\textwidth, height=0.7\textwidth]{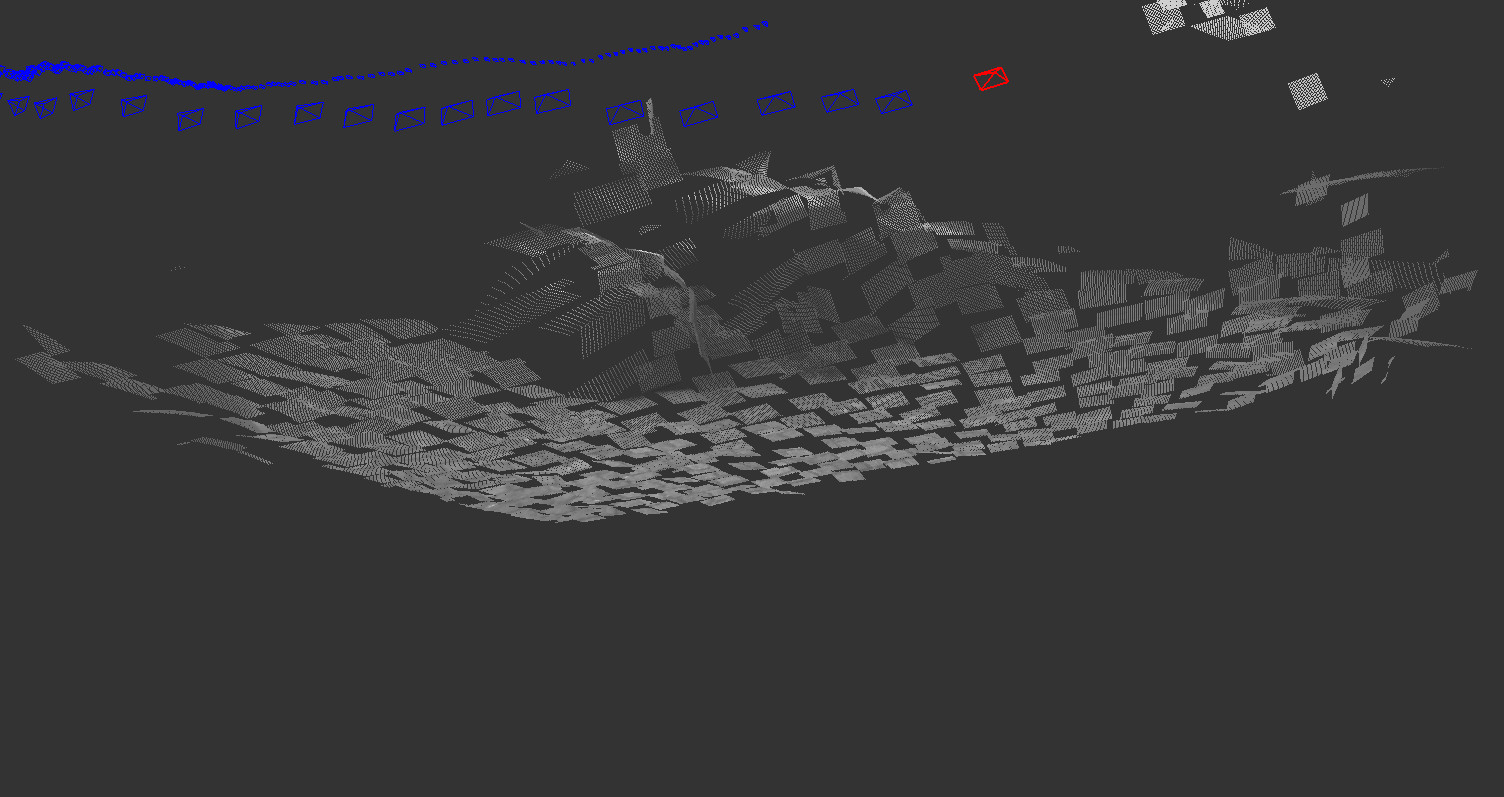}
     \end{subfigure}
     \begin{subfigure}[b]{0.24\textwidth}
         \centering
         \includegraphics[width=0.99\textwidth, height=0.7\textwidth]{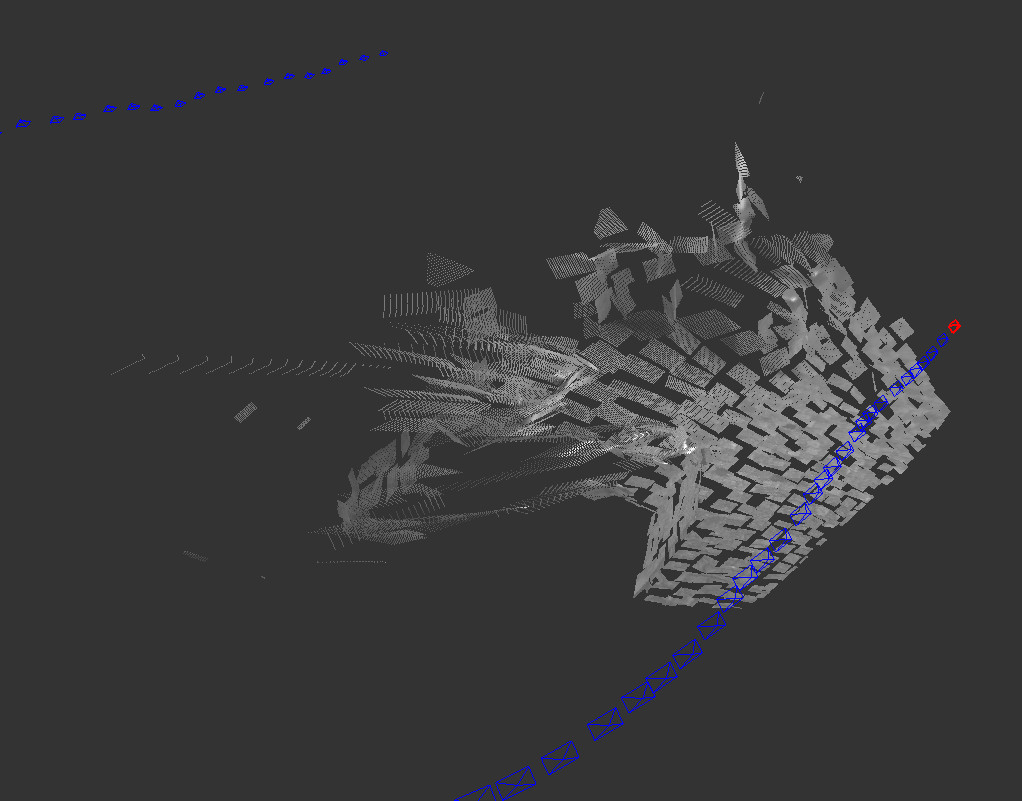}
     \end{subfigure}
     \begin{subfigure}[b]{0.24\textwidth}
         \centering
         \includegraphics[width=0.99\textwidth, height=0.7\textwidth]{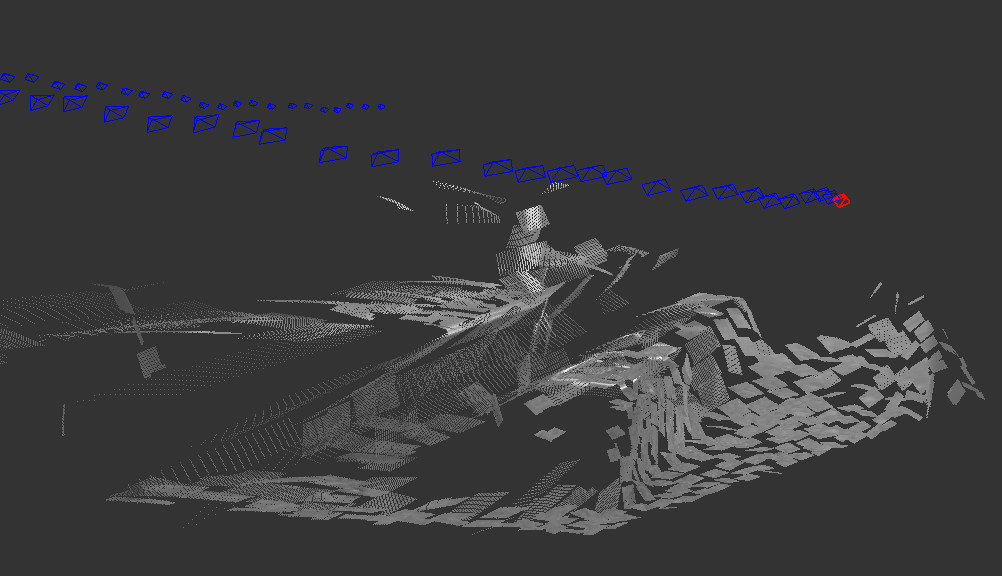}
     \end{subfigure}
     \begin{subfigure}[b]{0.24\textwidth}
         \centering
         \includegraphics[width=0.99\textwidth, height=0.7\textwidth]{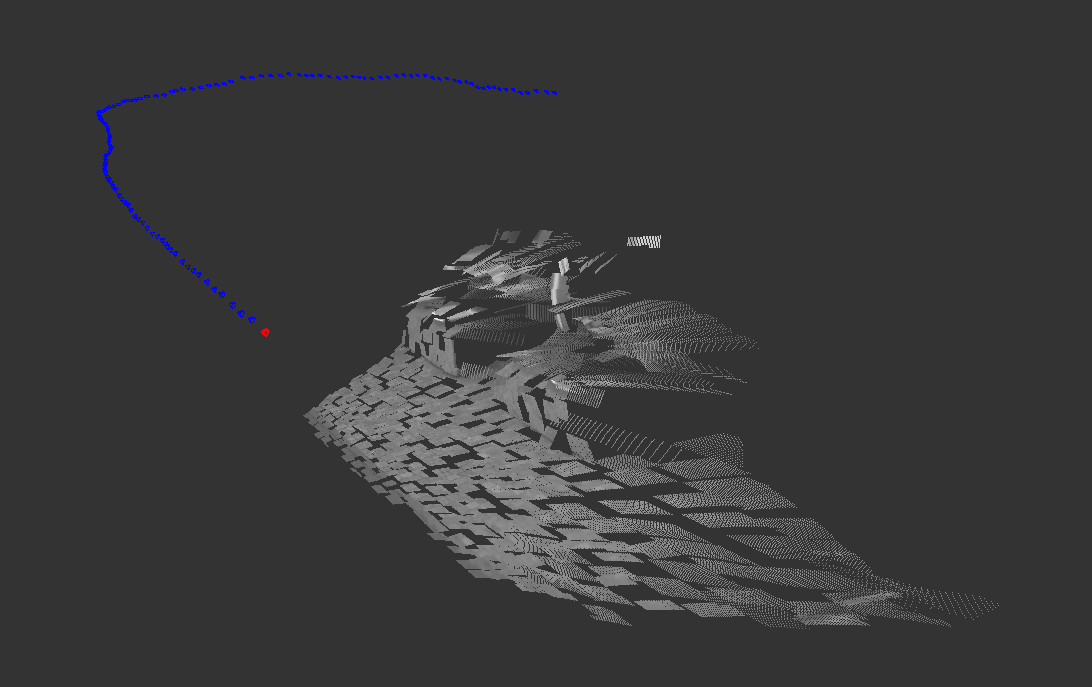}
     \end{subfigure}
     \begin{subfigure}[b]{0.24\textwidth}
         \centering
         \includegraphics[width=0.99\textwidth, height=0.7\textwidth]{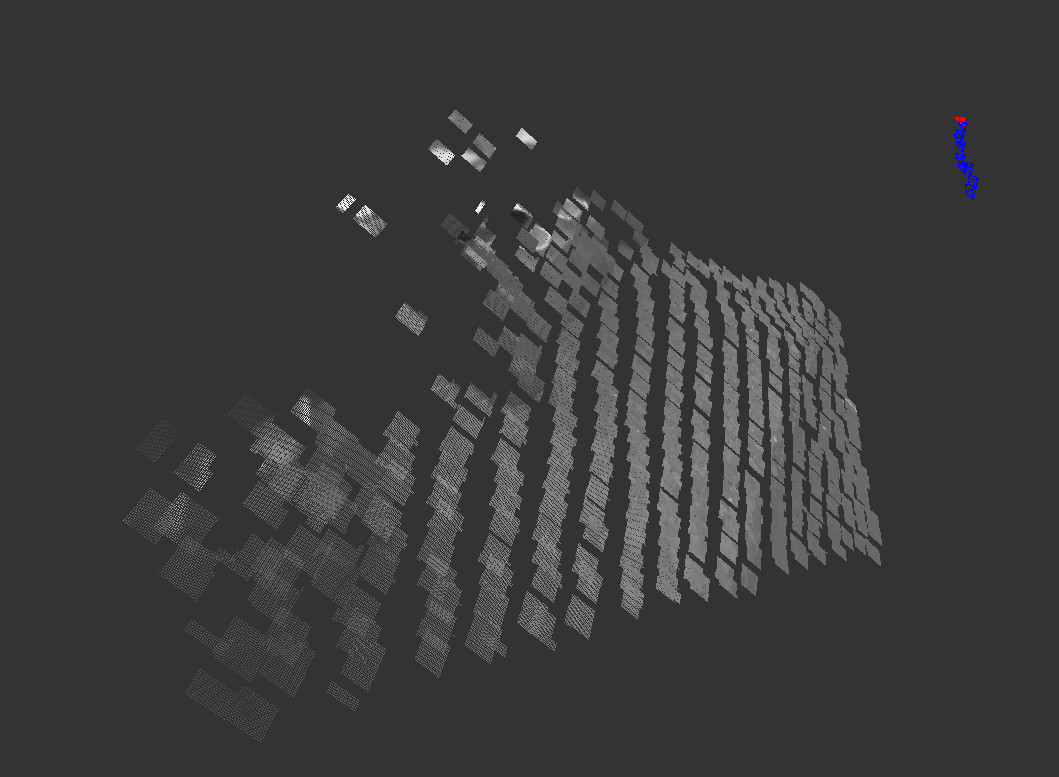}
     \end{subfigure}
     \begin{subfigure}[b]{0.24\textwidth}
         \centering
         \includegraphics[width=0.99\textwidth, height=0.7\textwidth]{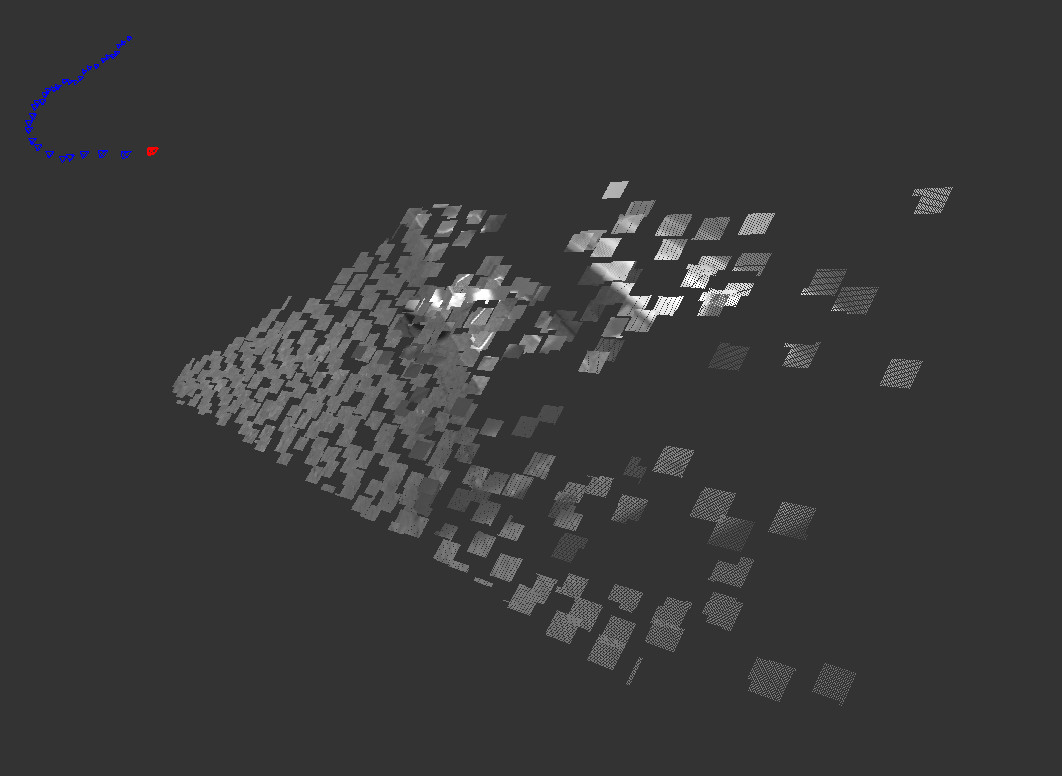}
     \end{subfigure}
     \begin{subfigure}[b]{0.24\textwidth}
         \centering
         \includegraphics[width=0.99\textwidth, height=0.7\textwidth]{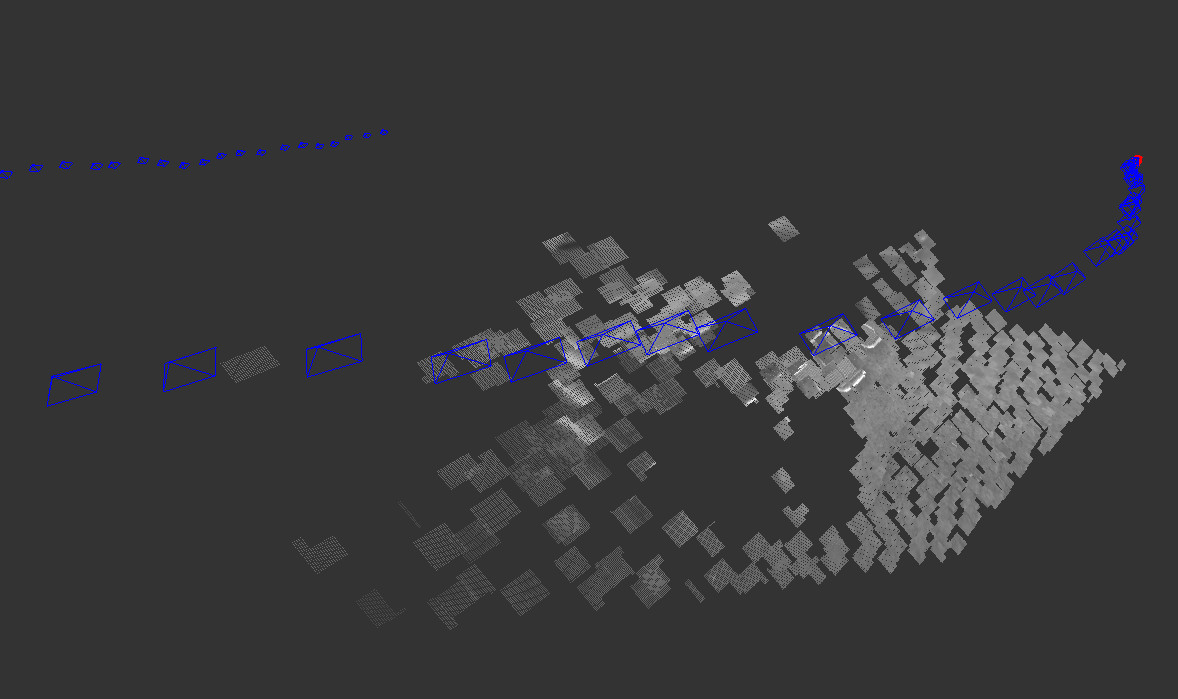}
     \end{subfigure}
     \begin{subfigure}[b]{0.24\textwidth}
         \centering
         \includegraphics[width=0.99\textwidth, height=0.7\textwidth]{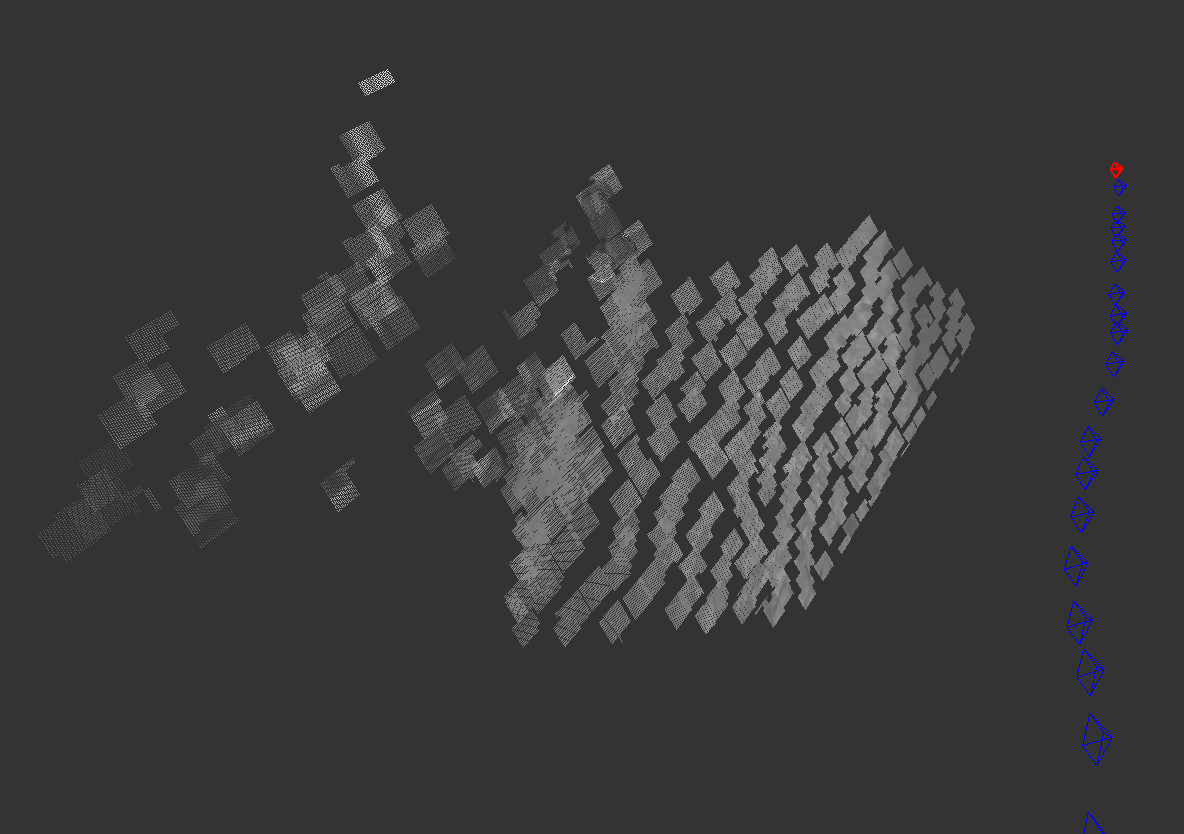}
     \end{subfigure}
     \caption{Detail of reconstruction with a surfel radius of 12 pixels. The first row shows result with the present method, while the second row show result without estimating the surfels normal. With the proposed method, even with such a coarse surfel reconstruction size, the benches and tables present in the scene can be still recognized, and the floor is correctly recovered. On the other hand, without estimating the surfels normal, the scene is almost completely unrecognizable, and there are heavy plane-depth related artifacts}
             \label{fig:surfel_12}
\end{figure}

In this evaluation, the surfel radius was set to 12 pixels wide. Results of the scene reconstruction can be seen in figure \ref{fig:surfel_12}. The widening of the surfel causes a coarse reconstruction result, loosing many details present on the scene. Nevertheless, the proposed approach is able to reconstruct the scene correctly, with the surfels on the floor and benches recovered correctly. Without using normal estimation the resulting reconstruction has a very poor quality, at the point of almost being unrecognizable, as seen in figure \ref{fig:surfel_12}.

\section{Conclusion}

The approach described here can estimate a monocular dense depth map as a set of surfels, directly from the raw image pixels. As far as the authors know, this is the first time this approach is utilized for monocular depth estimation. This presents several advantages over methods more commonly found in the literature, that implement pixel depth estimation followed by a regularization schema. Firstly, there is no need for a computationally costly and physically unlikely depth regularization. Also, the information from the normal estimation can be used to initialize new neighboring surfels, thus being able to initialize the Gauss Newton optimization approach without the need to perform further computations. 

The method was tested in several datasets, showing that the depth and surfels can be recovered correctly. The normal estimation allows for a more precise scene reconstruction, even when the surfel pixel radius is incremented considerably.

The main monocular depth estimation approach present in the literature revolves around pixelwise depth estimation in conjunction with depth regularization methods. The results obtained in this paper suggest that more plausible depth priors can be utilized for scene reconstruction.

\section*{References}

\bibliographystyle{splncs04}
\bibliography{Bibliography}

\begin{thebibliography}{10}
\providecommand{\url}[1]{\texttt{#1}}
\providecommand{\urlprefix}{URL }
\providecommand{\doi}[1]{https://doi.org/#1}

\bibitem{tum_lsd_slam_datasets}
Tum lsd-slam datasets. \url{https://vision.in.tum.de/research/vslam/lsdslam},
  accessed: 2017-07-22

\bibitem{Cadena_et_all_2016}
{Cadena}, C., {Carlone}, L., {Carrillo}, H., {Latif}, Y., {Scaramuzza}, D.,
  {Neira}, J., {Reid}, I., {Leonard}, J.J.: Past, present, and future of
  simultaneous localization and mapping: Toward the robust-perception age. IEEE
  Transactions on Robotics  \textbf{32}(6),  1309--1332 (Dec 2016).
  \doi{10.1109/TRO.2016.2624754}

\bibitem{Engel-et-al-pami2018}
Engel, J., Koltun, V., Cremers, D.: Direct sparse odometry. IEEE Transactions
  on Pattern Analysis and Machine Intelligence  (Mar 2018)

\bibitem{Engel_et_al_2014}
Engel, J., Sch\"ops, T., Cremers, D.: {LSD-SLAM}: Large-scale direct monocular
  {SLAM}. In: European Conference on Computer Vision (ECCV) (September 2014)

\bibitem{Goodfellow-et-al-2016}
Goodfellow, I., Bengio, Y., Courville, A.: Deep Learning. MIT Press (2016),
  \url{http://www.deeplearningbook.org}

\bibitem{Grisetti_et_all_2010}
{Grisetti}, G., {Kummerle}, R., {Stachniss}, C., {Burgard}, W.: A tutorial on
  graph-based slam. IEEE Intelligent Transportation Systems Magazine
  \textbf{2}(4),  31--43 (winter 2010). \doi{10.1109/MITS.2010.939925}

\bibitem{Hosni_et_al_2013}
{Hosni}, A., {Rhemann}, C., {Bleyer}, M., {Rother}, C., {Gelautz}, M.: Fast
  cost-volume filtering for visual correspondence and beyond. IEEE Transactions
  on Pattern Analysis and Machine Intelligence  \textbf{35}(2),  504--511 (Feb
  2013). \doi{10.1109/TPAMI.2012.156}

\bibitem{Mur-Artal_et_al_2015}
Mur-Artal, R., Montiel, J.M.M., Tardós, J.D.: Orb-slam: A versatile and
  accurate monocular slam system. IEEE Transactions on Robotics
  \textbf{31}(5),  1147--1163 (Oct 2015). \doi{10.1109/TRO.2015.2463671}

\bibitem{Newcombe_et_al_2011a}
Newcombe, R.A., Lovegrove, S.J., Davison, A.J.: Dtam: Dense tracking and
  mapping in real-time. In: 2011 International Conference on Computer Vision.
  pp. 2320--2327 (Nov 2011). \doi{10.1109/ICCV.2011.6126513}

\bibitem{Pizzoli_et_al_2014}
Pizzoli, M., Forster, C., Scaramuzza, D.: {REMODE}: Probabilistic, monocular
  dense reconstruction in real time. In: IEEE International Conference on
  Robotics and Automation (ICRA) (2014)

\bibitem{Wang_et_al_2019}
{Wang}, K., {Gao}, F., {Shen}, S.: Real-time scalable dense surfel mapping. In:
  2019 International Conference on Robotics and Automation (ICRA). pp.
  6919--6925 (May 2019). \doi{10.1109/ICRA.2019.8794101}

\bibitem{Whelan_et_al_2016}
Whelan, T., Salas-Moreno, R.F., Glocker, B., Davison, A.J., Leutenegger, S.:
  Elasticfusion: Real-time dense slam and light source estimation. The
  International Journal of Robotics Research  \textbf{35}(14),  1697--1716
  (2016). \doi{10.1177/0278364916669237},
  \url{https://doi.org/10.1177/0278364916669237}

\bibitem{Yan_et_al_2017}
{Yan}, Z., {Ye}, M., {Ren}, L.: Dense visual slam with probabilistic surfel
  map. IEEE Transactions on Visualization and Computer Graphics
  \textbf{23}(11),  2389--2398 (Nov 2017). \doi{10.1109/TVCG.2017.2734458}

\bibitem{Zienkiewicz_et_al_2016}
{Zienkiewicz}, J., {Tsiotsios}, A., {Davison}, A., {Leutenegger}, S.:
  Monocular, real-time surface reconstruction using dynamic level of detail.
  In: 2016 Fourth International Conference on 3D Vision (3DV). pp. 37--46 (Oct
  2016). \doi{10.1109/3DV.2016.82}

\end{thebibliography}

\end{document}